\documentclass{article} 

\usepackage{microtype}
\usepackage{graphicx}
\usepackage{booktabs} 

\usepackage{hyperref}

\usepackage[accepted]{icml_style/icml2024}

\usepackage{amsmath}
\usepackage{amssymb}
\usepackage{mathtools}
\usepackage{amsthm}

\usepackage[capitalize,noabbrev]{cleveref}

\theoremstyle{plain}

\theoremstyle{definition}

\theoremstyle{remark}

\usepackage[T1]{fontenc}
\usepackage{bm}
\usepackage{caption}
\usepackage{subcaption}
\usepackage{multirow}
\usepackage{array}
\usepackage[percent]{overpic}
\usepackage{listings}

\usepackage{pythonhighlight}
\usepackage{algorithm}
\usepackage{algorithmic}

\usepackage[disable]{todonotes} 

\usepackage{tikz}
\usetikzlibrary{positioning, arrows.meta, shadows.blur, backgrounds, calc}
\usepackage{xcolor}

\definecolor{mycolor}{rgb}{0.,0.2,0.7}
\definecolor{grey}{RGB}{169, 169, 169}
\definecolor{red}{RGB}{255, 0, 0}
\definecolor{blue}{RGB}{0, 0, 255}
\definecolor{orange}{RGB}{255, 165, 0}
\definecolor{purple}{RGB}{128, 0, 128}
\definecolor{skyblue}{RGB}{135, 206, 235}
\definecolor{confcolor}{RGB}{30, 150, 50} 

\def\smallnodedistance{0.45}
\def\smallnodesize{0.2}

\definecolor{grey}{RGB}{169, 169, 169}
\definecolor{red}{RGB}{255, 0, 0}

\newcommand{\atomembedding}[2]{ 
\pgfmathsetlengthmacro{\xshift}{-0.5*\smallnodedistance cm + #1 cm}
\begin{scope}[node distance=\smallnodedistance cm, yshift= #2 cm, xshift=\xshift]

    \tikzstyle{smallnode} = [circle, draw, minimum size=\smallnodesize cm]  

    \node[smallnode, fill=blue!70] (B) at (0,0) {}; 
    
    \node[smallnode, fill=green!15] (A) at ($(B)+(135:\smallnodedistance cm)$) {};
    \node[smallnode, fill=purple!70] (C) at ($(B)+(0:\smallnodedistance cm)$) {}; 
    \node[smallnode, fill=orange] (D) at ($(B)+(-90:\smallnodedistance cm)$) {}; 
    \node[smallnode, fill=green!50] (E) at ($(C)+(45:\smallnodedistance cm)$) {}; 

    \coordinate (I2) at ($(A)!0.5!(D) + (0.1*\smallnodedistance cm, 0.1*\smallnodedistance cm) $);

    \draw[black, thick, rotate around={-65:(I2)}] (I2) ellipse (1.9*\smallnodedistance cm and 0.9*\smallnodedistance cm);

\end{scope}
}

\newcommand{\atomembeddinggraph}[2]{ 
\pgfmathsetlengthmacro{\xshift}{-0.5*\smallnodedistance cm + #1 cm}
\begin{scope}[node distance=\smallnodedistance cm, yshift= #2 cm, xshift=\xshift]

    \tikzstyle{smallnode} = [circle, draw, minimum size=\smallnodesize cm]  

    \node[smallnode, fill=blue!70] (B) at (0,0) {}; 
    
    \node[smallnode, fill=green!15] (A) at ($(B)+(135:\smallnodedistance cm)$) {};
    \node[smallnode, fill=purple!70] (C) at ($(B)+(0:\smallnodedistance cm)$) {}; 
    \node[smallnode, fill=orange] (D) at ($(B)+(-90:\smallnodedistance cm)$) {}; 
    \node[smallnode, fill=green!50] (E) at ($(C)+(45:\smallnodedistance cm)$) {}; 

    \coordinate (I2) at ($(A)!0.5!(D) + (0.1*\smallnodedistance cm, 0.1*\smallnodedistance cm) $);

    \draw[black, thick, rotate around={-65:(I2)}] (I2) ellipse (1.9*\smallnodedistance cm and 0.9*\smallnodedistance cm);

    \draw (A) -- (B) -- (C) -- (E);
    \draw (B) -- (D);

\end{scope}
}

\newcommand{\anglegraphblack}[2]{ 
\pgfmathsetlengthmacro{\xshift}{-0.5*\smallnodedistance cm + #1 cm}
    \begin{scope}[node distance=\smallnodedistance cm, yshift= #2 cm, xshift=\xshift]

        \tikzstyle{smallnode} = [circle, draw, minimum size=\smallnodesize cm]  

        \node[smallnode, fill=white] (B) at (0,0) {}; 

        \node[smallnode, fill=white] (A) at ($(B)+(135:\smallnodedistance cm)$) {};

        \node[smallnode, fill=white] (D) at ($(B)+(-90:\smallnodedistance cm)$) {}; 
        
        \draw (A) -- (B) -- (D);

    \end{scope}
}

\newcommand{\torsiongraph}[2]{ 
\pgfmathsetlengthmacro{\xshift}{-0.5*\smallnodedistance cm + #1 cm}
    \begin{scope}[node distance=\smallnodedistance cm, yshift= #2 cm, xshift=\xshift]

        \tikzstyle{smallnode} = [circle, draw, minimum size=\smallnodesize cm]  
    
        \node[smallnode, fill=grey] (B) at (0,0) {}; 
        
        \node[smallnode, fill=grey!30] (A) at ($(B)+(135:\smallnodedistance cm)$) {};
        \node[smallnode, fill=grey] (C) at ($(B)+(0:\smallnodedistance cm)$) {}; 
        \node[smallnode, fill=red] (D) at ($(B)+(-90:\smallnodedistance cm)$) {}; 
        \node[smallnode, fill=grey!30] (E) at ($(C)+(45:\smallnodedistance cm)$) {}; 
        
        \coordinate (I3) at ($(B)!0.5!(C) + (.0cm, 0.5*\smallnodedistance cm) $);
    
        \draw (A) -- (B) -- (C) -- (E);
        \draw (B) -- (D);
    
        \draw[orange, thick, rotate around={-0:(I3)}] (I3) ellipse (2.2*\smallnodedistance cm and 1.1*\smallnodedistance cm);

    \end{scope}
}

\newcommand{\anglegraph}[2]{ 
\pgfmathsetlengthmacro{\xshift}{-0.5*\smallnodedistance cm + #1 cm}
    \begin{scope}[node distance=\smallnodedistance cm, yshift= #2 cm, xshift=\xshift]

        \tikzstyle{smallnode} = [circle, draw, minimum size=\smallnodesize cm]  

        \node[smallnode, fill=grey] (B) at (0,0) {}; 
        
        \node[smallnode, fill=grey!30] (A) at ($(B)+(135:\smallnodedistance cm)$) {};
        \node[smallnode, fill=grey] (C) at ($(B)+(0:\smallnodedistance cm)$) {}; 
        \node[smallnode, fill=red] (D) at ($(B)+(-90:\smallnodedistance cm)$) {}; 
        \node[smallnode, fill=grey!30] (E) at ($(C)+(45:\smallnodedistance cm)$) {}; 
    
        \coordinate (I2) at ($(A)!0.5!(D) + (0.1*\smallnodedistance cm, 0.1*\smallnodedistance cm) $);
        \draw (A) -- (B) -- (C) -- (E);
        \draw (B) -- (D);
    
        \draw[blue, thick, rotate around={-65:(I2)}] (I2) ellipse (1.9*\smallnodedistance cm and 0.9*\smallnodedistance cm);
    \end{scope}
}

\newcommand{\bondgraph}[2]{ 
\pgfmathsetlengthmacro{\xshift}{-0.5*\smallnodedistance cm + #1 cm}
    \begin{scope}[node distance=\smallnodedistance cm, yshift= #2 cm, xshift=\xshift]

        \tikzstyle{smallnode} = [circle, draw, minimum size=\smallnodesize cm]  

        \node[smallnode, fill=grey] (B) at (0,0) {}; 
        
        \node[smallnode, fill=grey!30] (A) at ($(B)+(135:\smallnodedistance cm)$) {};
        \node[smallnode, fill=grey] (C) at ($(B)+(0:\smallnodedistance cm)$) {}; 
        \node[smallnode, fill=red] (D) at ($(B)+(-90:\smallnodedistance cm)$) {}; 
        \node[smallnode, fill=grey!30] (E) at ($(C)+(45:\smallnodedistance cm)$) {}; 
        \coordinate (I) at ($(A)!0.5!(B)$);

        \draw (A) -- (B) -- (C) -- (E);
        \draw (B) -- (D);

        \draw[green, thick, rotate around={-45:(I)}] (I) ellipse (1.3*\smallnodedistance cm and 0.6*\smallnodedistance cm);
    \end{scope}
}

\newcommand{\moleculargraph}[2]{ 
    \pgfmathsetlengthmacro{\xshift}{-0.5*\smallnodedistance cm + #1 cm}

    \begin{scope}[node distance=\smallnodedistance cm, yshift= #2 cm, xshift=\xshift]
        \tikzstyle{smallnode} = [circle, draw, minimum size=\smallnodesize cm]  
    
        \node[smallnode, fill=grey] (B) at (0,0) {}; 
        
        \node[smallnode, fill=grey!30] (A) at ($(B)+(135:\smallnodedistance cm)$) {};
        \node[smallnode, fill=grey] (C) at ($(B)+(0:\smallnodedistance cm)$) {}; 
        \node[smallnode, fill=red] (D) at ($(B)+(-90:\smallnodedistance cm)$) {}; 
        \node[smallnode, fill=grey!30] (E) at ($(C)+(45:\smallnodedistance cm)$) {}; 
    
        \draw (A) -- (B) -- (C) -- (E);
        \draw (B) -- (D);
        
    \end{scope}
}

\begin{document}

\twocolumn[
\icmltitle{Grappa - A Machine Learned Molecular Mechanics Force Field
}



\icmlsetsymbol{equal}{*}

\begin{icmlauthorlist}
\icmlauthor{Leif Seute}{hits}
\icmlauthor{Eric Hartmann}{hits,iwr}
\icmlauthor{Jan St\"uhmer}{hits,kit}
\icmlauthor{Frauke Gr\"ater}{hits,iwr}
\end{icmlauthorlist}

\icmlaffiliation{hits}{Heidelberg Institute for Theoretical Studies, Schloss-Wolfsbrunnenweg 35, 69118 Heidelberg, Germany}
\icmlaffiliation{iwr}{Interdisciplinary Center for Scientific Computing, Heidelberg University, INF 205, 69120 Heidelberg, Germany}
\icmlaffiliation{kit}{Institute for Anthropomatics and Robotics, Karlsruhe Institute of Technology, Kaiserstr. 12, 76131 Karlsruhe, Germany}

\icmlcorrespondingauthor{Leif Seute}{leif.seute@h-its.org}

\icmlkeywords{Machine Learning, Force Field}

\vskip 0.3in
]



\printAffiliationsAndNotice{}  

\begin{abstract}
Simulating large molecular systems over long timescales requires force fields that are both accurate and efficient.
In recent years, E(3) equivariant neural networks have lifted the tension between computational efficiency and accuracy of force fields, but they are still several orders of magnitude more expensive than established molecular mechanics (MM) force fields.
Here, we propose Grappa, a machine learning framework to predict MM parameters from the molecular graph, employing a graph attentional neural network and a transformer with symmetry-preserving positional encoding.
The resulting Grappa force field outperformstabulated and machine-learned MM force fields in terms of accuracy at the same computational efficiency and can be used in existing Molecular Dynamics (MD) engines like GROMACS and OpenMM.
It predicts energies and forces of small molecules, peptides, RNA and --- showcasing its extensibility to uncharted regions of chemical space --- radicals at state-of-the-art MM accuracy. 
We demonstrate Grappa's transferability to macromolecules in MD simulations from a small fast folding protein up to a whole virus particle. Our force field sets the stage for biomolecular simulations closer to chemical accuracy, but with the same computational cost as established protein force fields.
\end{abstract}


\section{Introduction}

In recent years, advances in geometric deep learning have led to the development of highly accurate machine learned force fields, reshaping the field of computational chemistry and Molecular Dynamics (MD) simulations.
E(3) equivariant neural networks~\cite{allegro,mace,unke_protein_ff,painn} are capable of predicting energies and forces of small molecules to great accuracy with lower computational cost than quantum mechanical (QM) methods.
However, these models are several orders of magnitude more expensive than Molecular Mechanics (MM) force fields, which employ a simple physics-inspired functional form to parametrize the potential energy surface of a molecular system, hence trading off accuracy in favor of efficiency.
For MD simulations of large systems such as proteins and polynucleotides, MM force fields are well established and widely used.
Established MM force fields rely on lookup tables with a finite set of atom types characterized by chemical properties of the atom and its bonded neighbors for parameterization.
Recently, the Espaloma approach~\cite{espaloma_paper} has demonstrated that machine learning can be used to increase the accuracy of MM force fields by learning to assign parameters based on a graph representation of the molecule with chemical properties that rely on expert knowledge, such as orbital hybridization states or formal charge, as input features.

In this work, we propose a novel machine learning framework, Grappa (Graph Attentional Protein Parametrization), to learn MM parameters directly from the molecular graph, improving accuracy on a broad range of chemical space and eliminating the need for hand-crafted features.
Grappa employs a graph attentional neural network to construct atom embeddings capable of representing chemical environments based on the 2D molecular graph, followed by a transformer~\cite{transformer} with symmetry-preserving positional encoding.
Since MM parameters only have to be predicted once per molecule, Grappa can be incorporated into highly optimized MM engines such as GROMACS~\cite{gromacs} and OpenMM~\cite{openmm}.
This allows energy and force evaluations with the same computational cost as traditional force fields, at state-of-the-art MM accuracy.

We show that Grappa outperforms traditional MM force fields and the machine-learned MM force field Espaloma on the Espaloma dataset~\cite{espaloma0.3_paper}, which contains over 14,000 molecules and more than one million states, covering small molecules, peptides and RNA.
Since Grappa uses no hand-crafted chemical features, it can be extended to uncharted regions of chemical space, which we demonstrate on the example of peptide radicals.
Grappa is transferable to individual macromolecules and assemblies such as proteins and viruses, which exhibit similar dynamics as established force fields.
Starting from an unfolded initial state, MD simulations of small proteins parametrized by Grappa recover experimentally determined folding structures of small proteins, suggesting that Grappa captures the physics underlying protein folding.
We demonstrate the efficiency of Grappa, which is inhereted from MM, by simulating a system of one million atoms with the proposed force field on a single GPU, with a similar number of timesteps per second as a highly performant E(3) equivariant neural network~\cite{allegro} on over 4,000 GPUs.

\subsection{Molecular Mechanics}

In MM, the potential energy of a system with a given molecular graph is expressed as a sum of contributions from different interactions.
Bonded interactions are described by functions of E(3)-invariant internal coordinates such as the lengths $r_{ij}$ of bonds between two atoms, angles $\theta_{ijk}$ between three consecutive atoms and dihedrals $\phi_{ijkl}$ of two planes spanned by four atoms.
For the dihedrals, one considers interactions between four atoms that are either consecutively bonded (torsions) or where three atoms are bonded to a central atom (impropers), which do not reflect an independent degree of freedom but are used to maintain planarity of certain chemical groups.
One commonly uses harmonic potentials for bond stretching and angle bending and a periodic function for the dihedral potential. The potential energy of all interactions along bonds then is given by
\begin{align}
    E_{\text{bonded}}\left(\mathbf{x}\right) = & \sum_{(ij)\in \text{bonds}} k_{ij} (r_{ij} - r_{ij}^{(0)})^2 \nonumber \\
    + &\sum_{(ijk)\in \text{angles}} k_{ijk} (\theta_{ijk} - \theta_{ijk}^{(0)})^2 \nonumber \\
    + &\sum_{(ijkl)\in \text{dihedrals}} \sum_{n=1}^{n_{\text{periodicity}}} k_{ijkl} \cos(n \phi_{ijkl}),
\end{align}
with the equilibrium values (of bonds and angles) $r_{ij}^{(0)}$ and $\theta_{ijk}^{(0)}$, and the force constants $k_{ij}$, $k_{ijk}$ and $k_{ijkl}$, as the set of MM parameters, which we denote as
\begin{align}
\bm{\xi} \equiv \big\{\xi_{ij\ldots}^{(l)}|\:l\in\{\text{\small bonds}, \text{\small angles}, \text{\small torsions}, \text{\small impropers}\}\big\}.
\end{align}
For the periodic dihedral potential, a common choice is a Fourier series with the constraint that the dihedral potential is extremal at and symmetric around zero, which eliminates the need for sine terms.
Additionally, atom pairs that are not included in such N-body bonded interaction terms contribute to the potential energy through pairwise nonbonded interaction, typically described by Lennard-Jones and Coulomb potentials.

Traditional MM force fields define a finite set of atom types determined by hand-crafted rules, which are used to assign the free parameters $\{k_{ij}, r_{ij}^{(0)}, \ldots\}$ based on lookup tables for possible combinations of atom types.
In Grappa, we replace this scheme by learning the parameters from the molecular graph directly, which allows for a more flexible and transferable description of the potential energy surface.

A fundamental limitation of standard MM is the assumption of a constant molecular graph topology, which is enforced by the use of harmonic bond potentials. While this restricts accuracy and prohibits the description of bond-changing chemical reactions, the physical interpretability of the potential energy function ensures that simulated systems remain stable, even in states that are poorly described by the force field.
\section{Grappa}

\begin{figure}
    \centering
    \resizebox{0.95\columnwidth}{!}{\begin{tikzpicture}[
    node distance=0cm, 
    very thick,
    transformer_gnn_style/.style={
        rectangle, 
        rounded corners, 
        minimum width=2.7cm,
        minimum height=1.3cm,
        text centered, 
        draw=black, 
        top color=mycolor!30, 
        bottom color=mycolor!35,
        blur shadow={shadow blur steps=3}
    },
    other_style/.style={
        rectangle, 
        rounded corners, 
        minimum width=2.5cm, 
        minimum height=1.1cm, 
        text centered, 
        draw=black, 
        top color=black!7, 
        bottom color=black!10,
        blur shadow={shadow blur steps=3}
    },
    other_style_wide/.style={
        rectangle, 
        rounded corners, 
        minimum width=3cm, 
        minimum height=1.1cm, 
        text centered, 
        draw=black, 
        top color=black!7, 
        bottom color=black!10,
        blur shadow={shadow blur steps=3}
    },
    arrow_style/.style={-stealth, very thick, rounded corners=10mm}
]

    \def\yshiftgraphs{-4.7}
    \def\xshiftgraphs{3.6}

    \matrix (m) [column sep=0.9cm, row sep=0.6cm, align=center] {
        & \node [other_style_wide] (g) {Molecular Graph}; & \\
        & \node [transformer_gnn_style] (rep) {\textbf{Graph Attention}\\\textbf{Network}}; & \\
        & \node [other_style_wide] (e) {Atom Embeddings}; & \\
        \node [transformer_gnn_style] (bond) {\textbf{Symmetric}\\\textbf{Transformer}}; & \node [transformer_gnn_style] (angle) {\textbf{Symmetric}\\\textbf{Transformer}}; & \node [transformer_gnn_style] (torsion) {\textbf{Symmetric}\\\textbf{Transformer}}; \\
        \node [other_style] (p1) {Bond\\Parameters}; & \node [other_style] (p2) {Angle\\Parameters}; & \node [other_style] (p3) {Dihedral\\Parameters}; \\
        &&\\
        &&\\
    };

    \moleculargraph{-2.8}{4.35}

    \node[style=circle, draw=none] (I) [above of=g, yshift=0cm] {};

    \draw [arrow_style] (g) -- (rep);
    \draw [arrow_style] (rep) -- (e);
    \draw [arrow_style] (e) -| (bond);
    \draw [arrow_style] (e) -- (angle);
    \draw [arrow_style] (e) -| (torsion);
    \draw [arrow_style] (bond) -- (p1);
    \draw [arrow_style] (angle) -- (p2);
    \draw [arrow_style] (torsion) -- (p3);

    \bondgraph{-\xshiftgraphs}{\yshiftgraphs}
    \anglegraph{0}{\yshiftgraphs}
    \torsiongraph{\xshiftgraphs}{\yshiftgraphs}

\end{tikzpicture}}
    \caption{
        Grappa predicts MM parameters in two steps.
        First, atom embeddings are predicted from the molecular graph with a graph neural network.
        Then, transformers with symmetric positional encoding followed by permutation invariant pooling maps the embeddings to MM parameters with desired permutation symmetries.
        Once the MM parameters are predicted, the potential energy surface can be evaluated with MM-efficiency for different spatial conformations.
        }
        \label{fig:overview}
\end{figure}

Inspired by the atom typing with hand-crafted rules in traditional MM force fields and in analogy to \citealt{espaloma_paper}, Grappa first predicts $d$-dimensional atom embeddings,
\begin{align}
    \bm{\nu} = \{ \nu_i \in \mathbb{R}^d | i \in \mathcal{V} \},\label{eq:atom:embed}
\end{align}
which can represent local chemical environments that are encoded in the structure of the molecular graph $\mathcal{G}=(\mathcal{V},\mathcal{E})$, where the set of nodes $\mathcal{V}$ represents the atoms and the set of edges $\mathcal{E}$ represents the bonds.
In a second step (Figure~\ref{fig:overview}), for each interaction type $l$ MM parameters $\xi^{(l)}$ are predicted from the embeddings of the atoms involved in the respective energy contribution,
\begin{align}
    \xi^{(l)}_{ij\dots} = \psi^{(l)}\left(\nu_i, \nu_j, \ldots\right),
\end{align}
using a transformer $\psi^{(l)}$ that is invariant under certain permutations.
With the energy function of MM, the predicted parameter set $\bm{\xi}$ defines a potential energy surface, which can finally be evaluated for different spatial conformations $\mathbf{x}$ of the molecule,
\begin{align}
    \label{eq:mm_energy}
    E(\mathbf{x}) = E_\text{MM}\left(\mathbf{x}, \bm{\xi}\right).
\end{align}
Since the mapping from molecular graph to energy is differentiable with respect to the model parameters and spatial positions, it can be optimized on predicting QM energies and forces end-to-end, as visualized in Figure~\ref{fig:training}.
Notably, the machine learning model does not depend on the spatial conformation of the molecule, thus it has to be evaluated only once per molecule and the computational cost of each subsequent energy evaluation is given by the MM energy functional.

Grappa currently only predicts bonded MM parameters since we expect that nonbonded interactions are not covered sufficiently by the monomeric datasets used for training, rendering the nonbonded parameters underdetermined.
The nonbonded parameters are taken from established MM force fields that can reproduce solute interactions and melting points, which we expect to be strongly dependent on nonbonded interactions.

\subsection{Permutation symmetries in MM}

For the mapping from atom embeddings $\bm{\nu}$ to MM parameters, we postulate certain permutation symmetries that the model should respect.
To derive these symmetries, we consider the energy function of MM as a decomposition into contributions from subgraphs of the featurized molecular graph that correspond to bonds, angles, torsions and improper dihedrals.
We demand invariance of the energy contribution under node permutations that induce isomorphisms of the respective subgraph.
For bonds, angles and torsions, these permutations leave the respective spatial coordinate invariant, thus we can achieve invariance of the energy contribution by demanding invariance of the MM parameters,
\begin{align}
\xi^{(\text{bond})}_{ij} &= \xi^{(\text{bond})}_{ji}\,, \label{eq:bond_symm}\\
\xi^{(\text{angle})}_{ijk} &= \xi^{(\text{angle})}_{kji}\,, \label{eq:angle_symm}\\
\xi^{(\text{torsion})}_{ijkl} &= \xi^{(\text{torsion})}_{lkji}\,. \label{eq:torsion_symm}
\end{align}
For improper dihedrals, however, not all subgraph isomorphisms leave the dihedral angle invariant and demanding parameter invariance under those permutations would lead to an energy contribution that is not invariant.
In Grappa, we solve this problem by decomposing the improper torsion contributions into three terms, as described in~\ref{sec:impropers}.

\subsection{The Grappa architecture}\label{sec:grappa}

\begin{figure}[h!]
    \centering

    \resizebox{0.99\columnwidth}{!}{\def\nodedistance{0.6}
\def\halfnodedistance{0.3}

\def\graphshiftx{3.3}
\def\graphshifty{-15}
\def\flowshifty{-7.5}

\def\shiftx{5.7}
\def\shifty{-2.1}

\def\totalshiftx{0.9}
\def\totalshiftxx{-\graphshiftx - \shiftx}

\begin{tikzpicture}[
    node distance=\nodedistance cm and 1.0cm,
    very thick,
    process/.style={
        rectangle,
        rounded corners,
        minimum width=3cm,
        minimum height=1cm,
        text centered,
        draw=black,
        top color=mycolor!30,
        bottom color=mycolor!35,
        blur shadow={shadow blur steps=3}
    },
    ln/.style={
        rectangle,
        rounded corners,
        minimum width=3cm,
        minimum height=1cm,
        text centered,
        draw=black,
        top color=mycolor!20,
        bottom color=mycolor!25,
        blur shadow={shadow blur steps=3}
    },
    other_style/.style={
        rectangle, 
        rounded corners, 
        minimum width=3cm, 
        minimum height=1cm, 
        text centered, 
        draw=black, 
        top color=black!7, 
        bottom color=black!10,
        blur shadow={shadow blur steps=3}
    },
    arrow/.style={
        ->,
        >=stealth,
        very thick,
        rounded corners
    },
    arrow_nohead/.style={
        -,
        very thick,
        rounded corners
    },
    plus/.style={
        circle,
        fill=white,
        draw,
        inner sep=1pt
    },
    smallnode/.style={
        circle,
        draw,
        minimum size=0.4cm,
        fill=grey!30
    },
    smallnode_highlight/.style={
        smallnode,
        fill=red
    },
    edge/.style={
        very thick,
        rounded corners
    },
    linear_style/.style={
        process, fill=blue!30
    },
]

\tikzstyle{edge} = [very thick,-, rounded corners]
\tikzstyle{plus} = [circle, fill=white, draw, inner sep=1pt]



\atomembedding{\totalshiftxx}{\shifty}

\anglegraphblack{\totalshiftx}{\shifty}


\begin{scope}[yshift=\flowshifty cm]

    \matrix (m) [column sep=2.5cm, row sep=\nodedistance cm, align=center] {
         \node [other_style, minimum height=1.2cm] (atoms) {Tuple of\\Atom Embeddings}; & \node [other_style] (subgraph) {Interaction Subgraph};\\
         \node [process] (lin1) {Nodewise Linear}; & \node [other_style, top color=purple!15, bottom color=purple!20, minimum width=3.8cm, minimum height=1.2cm] (encoding) {Symmetric\\Positional Encoding};\\
         \node [ln] (ln1) {Layer Norm}; & \node [ln] (permute) {Permute, Concat};\\
         \node [process, minimum height=1.2cm] (att) {\textbf{Multihead}\\ \textbf{Attention}}; & \node [process] (ff_symm) {\textbf{Feed-Forward}};\\
         \node [ln] (ln2) {Layer Norm}; & \node [ln] (symm) {Sum};\\
         \node [process, minimum height=1.2cm] (ff) {\textbf{Nodewise}\\ \textbf{Feed-Forward}}; & \node [ln] (interval) {To Interval};\\
         & \node [other_style] (param) {Interaction Parameters};\\
    };

    \draw [arrow] (atoms) -- (lin1);
    \draw [edge] (subgraph) -- (encoding);
    \draw [arrow] (encoding) -- (lin1);

    \draw [arrow] ([yshift=0 cm]ff.south) |- ++(3.6,-1.3) |-  ([yshift=-0 cm]permute.west);

    \draw [arrow] (lin1) -- (ln1);
    \draw [arrow] (ln1) -- (att);
    \draw [edge] (att) -- (ln2);
    \draw [arrow] (ln2) -- (ff);
    \draw [arrow] (permute) -- (ff_symm);
    \draw [arrow] (ff_symm) -- (symm);
    \draw [arrow] (symm) -- (interval);
    \draw [arrow] (interval) -- (param);

    \draw [edge] ([yshift=\halfnodedistance cm]att.north) -- ++(2.3,0) |- ([yshift=-\halfnodedistance cm]att.south);
    \draw [edge] ([yshift=\halfnodedistance cm]ff.north) -- ++(2.3,0) |- ([yshift=-\halfnodedistance cm]ff.south);

    \node [plus] at ([yshift=-\halfnodedistance cm]att.south) {+};
    \node [plus] at ([yshift=-\halfnodedistance cm]ff.south) {+};

\end{scope}

\begin{scope}[on background layer]
    \draw[rounded corners, thin, draw=black, fill=black!11] ([shift={(-1.3cm,\halfnodedistance cm)}]ln1.north west) rectangle ([shift={(1.3cm,-\nodedistance cm)}]ff.south east) coordinate (rect_se) at (current path bounding box.south east);
\end{scope}


\node [left=0.5cm of ln1, anchor=west, scale=1.5, yshift=0.35cm, xshift=-0.55cm] {\(n \times \)};

\end{tikzpicture}}
    \caption{
        Architecture of the symmetric transformer:
        Atom embeddings are equipped with a permutation invariant positional encoding determined by the subgraph they represent.
        They are then passed through $n=3$ permutation equivariant transformer layers, symmetry-pooled and mapped to the possible range of the respective parameter.
        }
        \label{fig:symmetric_transformer}
\end{figure}

To predict atom embeddings from the molecular graph, Grappa employs a graph attentional neural network inspired by the transformer architecture~\cite{transformer}.
Multi-head dot-product attention on graph edges~\cite{gat_original} is followed by feed-forward layers with residual connections~\cite{resnet} and layer normalization~\cite{layernorm}, which has been demonstrated to enhance the expressivity of attention layers~\cite{brody2023expressivity}.

For the map from these embeddings to MM parameters, it is desirable to use an architecture that respects the permutation symmetries (Eqs. \ref{eq:bond_symm} - \ref{eq:torsion_symm}) by design, constraining the space of possible models to those that are physically sensible.
In the spirit of equivariant machine learning, we use permutation equivariant layers followed by final symmetric pooling.
However, since we do not require invariance under all permutations but only under permutations as defined in Eqs. \ref{eq:bond_symm} - \ref{eq:torsion_symm}, we can increase expressivity by allowing the model to break symmetries that are not required.

Following these considerations, we use a transformer architecture based on multi-head dot-attention with a positional encoding that is invariant under the required symmetries but can break others.
For example for angles, this positional encoding is given by
\begin{align}
    (\nu_i, \nu_j, \nu_k) \mapsto \left(\nu_i\oplus0,\: \nu_j\oplus1, \: \nu_k\oplus0\right), \label{eq:pos_enc}
\end{align}
where the $\oplus$ operation appends the respective value to the node feature vector, making it invariant under $ijk\rightarrow kji$ but not under e.g. $ijk\rightarrow jik$.
After these equivariant layers, we apply a multilayer perceptron (MLP) on the concatenated permuted node embeddings and sum over the desired set of permutations $\mathcal{P}$,
\begin{align}
    z_{ij\dots} = \sum_{\sigma\in \mathcal{P}} \text{MLP}\left(\left[\nu_{\sigma(i)}, \nu_{\sigma(j)}, \dots\right]\right), \label{eq:pooling}
\end{align}
defining a symmetry pooling operation with $\mathcal{P}$-invariant output.
We call this combination of permutation invariant positional encoding with permutation equivariant layers and symmetric pooling the \textit{symmetric transformer}, which is illustrated in Figure~\ref{fig:symmetric_transformer}.
The symmetric transformer can be generalized to permutation symmetries of arbitrary subgraphs by using the eigenvectors of the graph Laplacian~\cite{graph_transformer} as positional encoding.

Finally, we map to the range of physically sensible parameters, e.g. $(0,\infty)$ for bond and angle force constants or $(0,\pi)$ for equilibrium angles $\theta^{(0)}$.
To this end, we use scaled and shifted versions of ELU and the sigmoid function as described in~\ref{sec:scaling}.
While one could also use the exponential for mapping to $(0,\infty)$, ELU's linear behaviour towards large inputs is favorable for producing stable gradients during optimization.
With the scaling we ensure that a normally distributed output of the neural network is mapped to a distribution with mean and standard deviation that is suitable for the respective MM parameter, which can be seen as a normalization technique~\cite{layernorm}.
For predicting dihedral parameters $\xi_{ijkl}$, we use a sigmoid gate, which allows the model to supress dihedral modes that are not needed.

As input feature, we use one-hot encodings of the atomic number, the number of neighbors of the respective node and membership in loops of length 3 to 8, which can be directly calculated from the molecular graph.
The assignment of nonbonded MM parameters is done using a traditional force field of choice.
Since the bonded parameters predicted by Grappa may depend on the scheme by which nonbonded parameters are assigned, we pass the partial charge of each atom as input feature to Grappa, which also allows to encode the total charge of a molecule without breaking graph symmetries as one would potentially do by using the formal charge instead.

\subsection{Training}

\begin{figure}
    \centering
    \resizebox{0.95\columnwidth}{!}{\usetikzlibrary{fit, backgrounds, positioning}

\begin{tikzpicture}[
    node distance=0cm,
    very thick,
]

\tikzset{
    bluestyle/.style={
        rectangle,
        rounded corners,
        minimum width=2.7cm,
        minimum height=1.3cm,
        text centered,
        draw=black,
        top color=mycolor!30,
        bottom color=mycolor!35,
        blur shadow={shadow blur steps=3}
    },
    other_style/.style={
        rectangle,
        rounded corners,
        minimum width=2.5cm,
        minimum height=1.1cm,
        text centered,
        draw=black,
        top color=black!7,
        bottom color=black!10,
        blur shadow={shadow blur steps=3}
    },
    other_style_wide/.style={
        rectangle,
        rounded corners,
        minimum width=3cm,
        minimum height=1.1cm,
        text centered,
        draw=black,
        top color=black!7,
        bottom color=black!10,
        blur shadow={shadow blur steps=3}
    },
    conf_style/.style={
        rectangle,
        rounded corners,
        minimum width=2.5cm,
        minimum height=1.1cm,
        text centered,
        draw=black,
        top color=confcolor!35,
        bottom color=confcolor!40,
        blur shadow={shadow blur steps=3}
    },
    ref_style/.style={
        rectangle,
        rounded corners,
        minimum width=2.5cm,
        minimum height=1.1cm,
        text centered,
        draw=black,
        top color=confcolor!25,
        bottom color=confcolor!30,
        blur shadow={shadow blur steps=3}
    },
    arrow_style/.style={
        -stealth,
        very thick,
        rounded corners=10mm
    },
    line_style/.style={
        very thick, rounded corners=10mm  
    }
}

\node at (5cm,1cm) {}; 

    \matrix (m) [column sep=0.9cm, row sep=0.6cm, align=center, yshift=-2.3cm] {
        \node[other_style, alias=mgraph] {Molecular Graph}; &
        \node[conf_style, alias=xyz, opacity=0, shadow scale=0] {Positions}; &
        \node[ref_style, alias=qm, opacity=0, shadow scale=0] {QM Energy\\QM Force}; \\
        \node[bluestyle, alias=grappa] {\textbf{Grappa}}; &
        & \\
        \node[other_style, alias=mmparams] {MM Parameters}; &
        & \\
        \node[bluestyle, alias=mmenergy] {\textbf{MM Energy}\\\textbf{Functional}}; & & \\
        \node[conf_style, alias=output, opacity=0, shadow scale=0] {Pred. Energy\\Pred. Force}; & &
        \node[ref_style, alias=loss] {Loss}; \\
    };

    \draw[arrow_style] (qm) -- (loss);
    
    \node[conf_style, opacity=0.4, shadow scale=0] at ([xshift=1mm, yshift=1mm] xyz) {};
    \node[conf_style, opacity=0.2, shadow scale=0] at ([xshift=2mm, yshift=2mm] xyz) {};
    \node[conf_style] at ([xshift=0mm, yshift=0mm] xyz) {Positions};

    \node[conf_style, opacity=0.4, shadow scale=0] at ([xshift=1mm, yshift=1mm] output) {};
    \node[conf_style, opacity=0.2, shadow scale=0] at ([xshift=2mm, yshift=2mm] output) {};
    \node[conf_style, align=left] at ([xshift=0mm, yshift=0mm] output) {Pred. Energy\\Pred. Force};

    \node[ref_style, opacity=0.4, shadow scale=0] at ([xshift=1mm, yshift=1mm] qm) {};
    \node[ref_style, opacity=0.2, shadow scale=0] at ([xshift=2mm, yshift=2mm] qm) {};
    \node[ref_style, align=left] at ([xshift=0mm, yshift=0mm] qm) {QM Energy\\QM Force};

    \draw[arrow_style] (mgraph) -- (grappa);
    \draw[arrow_style] (grappa) -- (mmparams);
    \draw[arrow_style] (mmparams) -- (mmenergy);
    \draw[arrow_style] (xyz) |- (mmenergy.east);
    \draw[arrow_style] (mmenergy) -- (output);
    \draw[arrow_style] (output) -- (loss);

    \draw[arrow_style, dashed, opacity=0.7] ([xshift=-0.5cm]loss.north) to[out=90, in=0] (grappa.east);

    \node at ($(loss.north)!0.5!(grappa.east) + (1.1cm, 1.15cm)$) [font=\itshape, rotate=-45, opacity=0.7] {\textit{Backprop.}};

    \begin{scope}[on background layer]
    \draw[rounded corners, thin, draw=black, fill=black!11] ([shift={(-0.5cm,0.7 cm)}]mgraph.north west) rectangle ([shift={(0.5cm,-0.3cm)}]mmparams.south east) coordinate (rect_se) at (current path bounding box.south east);
    \end{scope}
    
    \node [left=-0.2cm of mgraph, anchor=west, yshift=1.0cm, xshift=-0.68cm, font=\itshape] {Inference};

\end{tikzpicture}}
    \caption{
        Grappa predicts one set of parameters per molecule.
        With the MM energy functional (Eq. \ref{eq:mm_energy}), the parameters can be mapped to energies and forces of given states, whose deviation from the ground truth is minimized during training.
        State-specific quantities are represented in green, molecule-specific quantities are represented in grey.
        }
        \label{fig:training}
\end{figure}

We train Grappa to minimize the mean squared error (MSE) between QM energies $E_{\text{QM}}$ and forces $-\nabla_\mathbf{x} E_\text{QM}$, whose contribution we weight by the hyperparameter $\lambda_F$, and its prediction, $E$ and $\nabla_\mathbf{x} E$ respectively, which includes the non-learnable nonbonded contribution.
At the start of training, we also include the deviation of predicted bond, angle and torsion MM parameters to those of a given traditional force field, weighted by $\lambda_{\text{trad}}$ and, for regularization, the $L2$ norm of dihedral MM parameters $\xi_{ijkl}^{(\text{dih})}$, weighted by $\lambda_{\text{dih}}$ as in~\cite{espaloma0.3_paper}. That is, we use the loss function
\begin{align}
    \mathcal{L} = \: &\text{MSE}\left(E, E_\text{QM}\right) + \lambda_F \text{MSE}\left(\nabla_\mathbf{x} E, \nabla_\mathbf{x} E_\text{QM}\right) \nonumber \\
    + &\lambda_{\text{MM}} \text{MSE}\left(\xi_\text, \xi_\text{trad}\right) + \lambda_{\text{dih}} \big\Vert\xi^{(\text{dih})}\big\Vert_2^2\,.
\end{align}
Since MM can only predict energy differences of states, not formation energies, we subtract the mean of target and predicted energies for each molecule.
As shown previously for other machine learned force fields~\cite{schnet, painn}, we find training on forces $-\nabla_r E$ in addition to energies important also in our setting, for two reasons.
First, the energy is a global, pooled quantity and thus less expressive than the forces, which are local and by a factor of $3N$ more numerous.
Second, as shown in~\cite{sobolev_training}, learning derivatives of the target with respect to the input can lead to improved generalization and data efficiency, effectively smoothing the learned potential energy surface.

Details on the training procedure and hyperparameters can be found in~\ref{sec:hyperparams}.
\section{Results}\label{sec:results}

\subsection{Grappa is state-of-the-art}\label{sec:benchmark}
To demonstrate that Grappa is the current state-of-the-art MM force field in terms of accuracy, we train and evaluate a Grappa model on the dataset reported in the follow-up paper~\cite{espaloma0.3_paper} to Espaloma, which contains 17,427 unique molecules and over one million conformations.
Our training and test partition is identical with the one from Espaloma, where the molecules were divided into 80\% training, 10\% validation and 10\% test set, based on isomeric SMILES strings.
The dataset covers small molecules, peptides and RNA with states sampled from the Boltzmann distribution at 300\,K and 500\,K, from optimization trajectories and from torsion scans.
The nonbonded contribution is calculated using the OpenFF-2.0.0 force field~\cite{openff2_paper} and partial charges from the AM1-BCC method, as in Espaloma.
We train the model for 1,000 epochs on an A100 GPU, which takes about one day.

\begin{figure*}[h!]
    \def\scatterplotratio{0.71}
    \def\pcaplotratio{0.23}
    \centering
    \begin{minipage}[t]{\scatterplotratio\textwidth}
        \includegraphics[width=\textwidth]{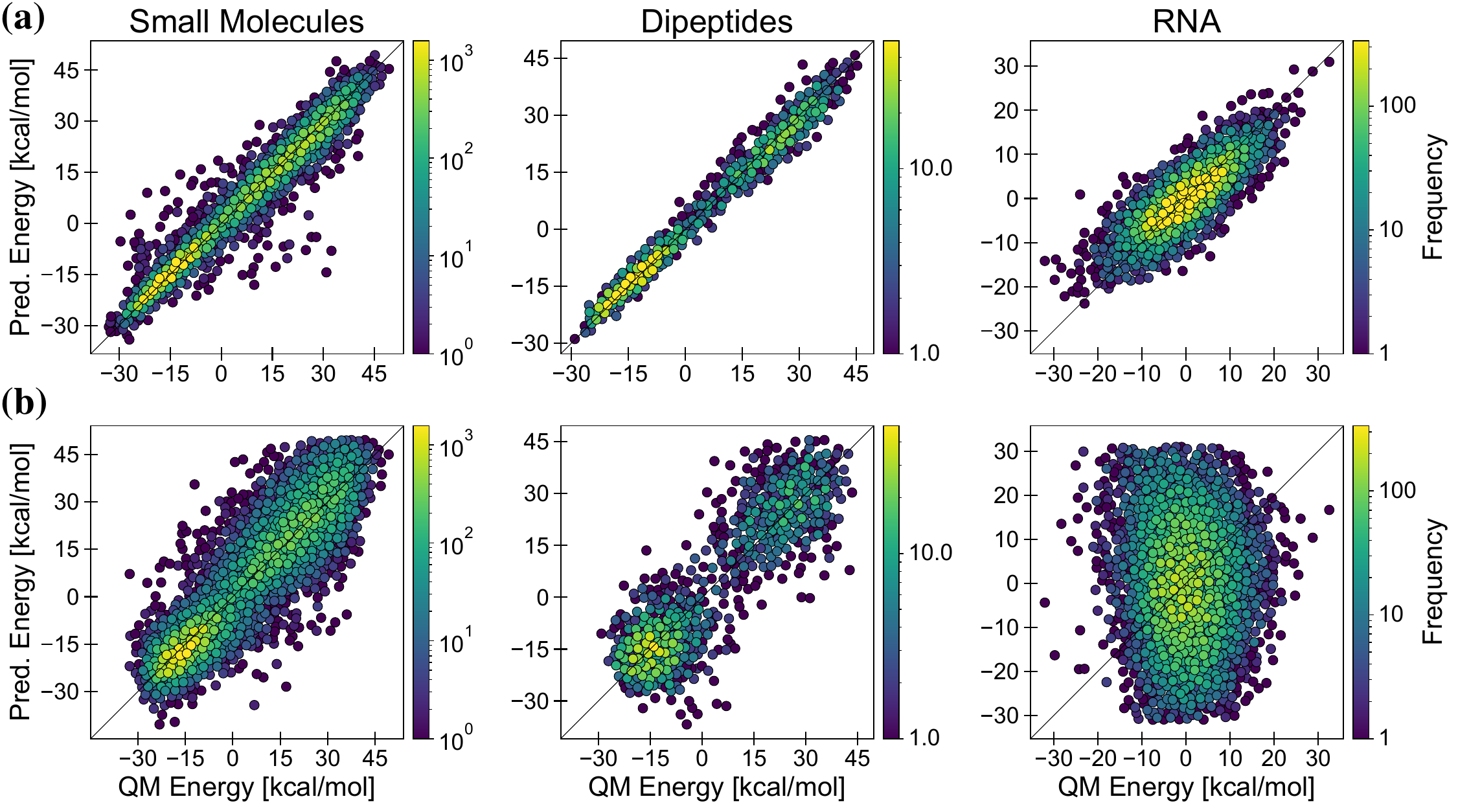}
    \end{minipage}%
    \quad\quad
    \begin{minipage}[t]{\pcaplotratio\textwidth}
        \includegraphics[width=1.05\textwidth]{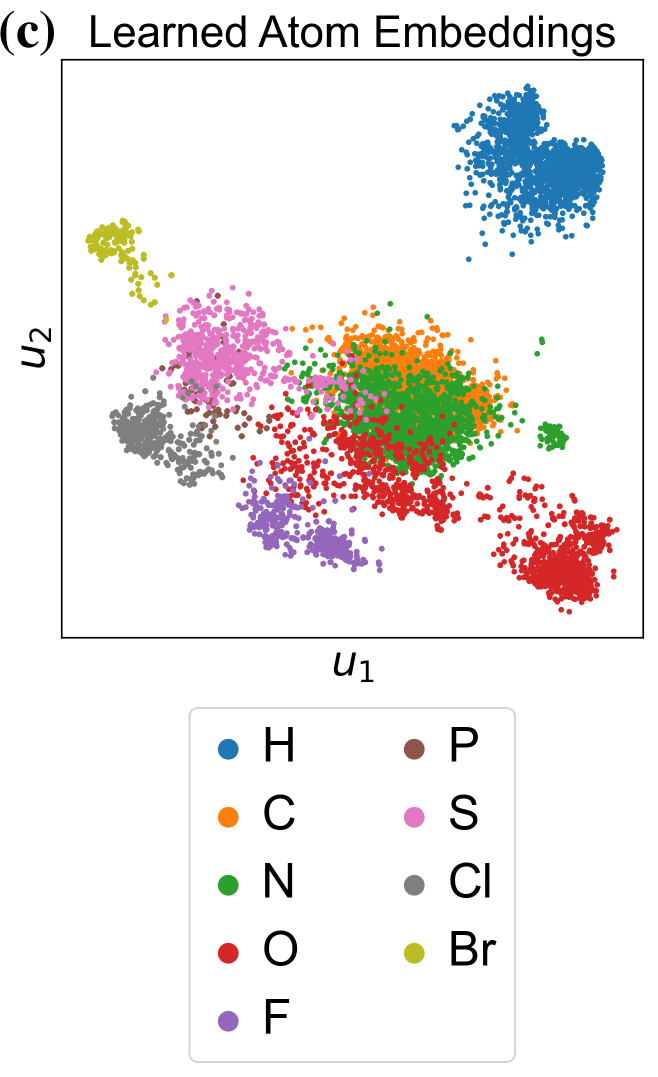}
    \end{minipage}
    \caption{
    Comparison of energy predictions of \textbf{(a)} \textbf{Grappa-1.3} and the established force fields \textbf{(b)} \textbf{Gaff-2.11}, \textbf{ff99SB-ILDN} and \textbf{RNA.OL3} for test molecules from Espaloma's SPICE-Pubchem, SPICE-Dipeptide and RNA-Trinucleotide datasets; force predictions are depicted at~\ref{fig:gradient_scatter}.
    \textbf{(c)} The first principal components $u_1$ and $u_2$ of predicted atom embeddings from the Espaloma test dataset can be related to a combination of the main group and period in the periodic table of elements. Lines of constant main group or period are represented by approximate diagonals in latent space.
    }
    \label{fig:scatter_and_pca}
\end{figure*}

For all types of molecules, Grappa outperforms established MM force fields and Espaloma in terms of energy and force accuracy on Boltzmann-sampled states as shown in Table~\ref{tab:benchmark} and Figure~\ref{fig:scatter_and_pca}a.
While the Boltzmann samples are the more relevant benchmarking data for using Grappa in MD simulations, we also compare performance on torsion scans and optimization trajectories (Table~\ref{tab:full_benchmark}). 
There, we find Grappa and Espaloma to be competitive, while Grappa is more accurate for forces and less accurate for energies.
Grappa also outperforms the baselines if only trained on a fraction of Espaloma's training set (Figure~\ref{fig:learning_curve}), indicating high data efficiency.
Chemical properties based on expert knowledge, e.g. hybridization and aromaticity, have long been used to assign MM parameters~\cite{gaff_paper,openff2_paper}. We show that accurate MM parameters can be predicted directly from the molecular graph, without relying on hand-crafted input features.

\begin{table*}[h!]
    
\newcommand{\widthbetweentype}{7pt}

\begin{tabular*}{\textwidth}{l c c l c c c c c c}

        \hline
        \hline
        \multirow{2}{*}{Dataset} & \multirow{2}{*}{Test Mols} & \multirow{2}{*}{Confs} & & \multirow{2}{*}{Grappa} & \multirow{2}{*}{Espaloma} & \multirow{2}{*}{Gaff-2.11} & ff14SB, & Mean\\
        & & & & & & & RNA.OL3 & Predictor\\
        \hline
        \multirow{2}{*}{SPICE-Pubchem} & \multirow{2}{*}{1411} & \multirow{2}{*}{60853} & \textit{Energy} & \textbf{2.3} & \textbf{2.3} & 4.6 &  & 18.4\\
                                           &                       &                         & \textit{Force}  & \textbf{6.1} & 6.8 & 14.6 &  & 23.4\\
        \hline
        \multirow{2}{*}{SPICE-DES-Monomers} & \multirow{2}{*}{39} & \multirow{2}{*}{2032} & \textit{Energy} & \textbf{1.3} & 1.4 & 2.5 &  & 8.2\\
                                           &                       &                         & \textit{Force}  & \textbf{5.2} & 5.9 & 11.1 &  & 21.3\\
        \hline
        \multirow{2}{*}{SPICE-Dipeptide} & \multirow{2}{*}{67} & \multirow{2}{*}{2592} & \textit{Energy} & \textbf{2.3} & 3.1 & 4.5 & 4.6 & 18.7\\
                                           &                       &                         & \textit{Force}  & \textbf{5.4} & 7.8 & 12.9 & 12.1 & 21.6\\
        \hline
        \multirow{2}{*}{RNA-Diverse} & \multirow{2}{*}{6} & \multirow{2}{*}{357} & \textit{Energy} & \textbf{3.3} & 4.2 & 6.5 & 6.0 & 5.4\\
                                           &                       &                         & \textit{Force}  & \textbf{3.7} & 4.4 & 16.7 & 19.4 & 17.1\\
        \hline
        \multirow{2}{*}{RNA-Trinucleotide} & \multirow{2}{*}{64} & \multirow{2}{*}{35811} & \textit{Energy} & \textbf{3.5} & 3.8 & 5.9 & 6.1 & 5.3\\
                                           &                       &                         & \textit{Force}  & \textbf{3.6} & 4.3 & 17.1 & 19.7 & 17.7\\
        \hline
        \hline
        \hline
\end{tabular*}

    \caption{
        Accuracy of Grappa, Espaloma~0.3~\cite{espaloma0.3_paper} and established MM force fields on test molecules of the Espaloma dataset.
        We report the RMSE of centered energies in kcal/mol and the componentwise RMSE of forces in kcal/mol/Å, uncertainties can be found in Table~\ref{tab:full_benchmark}.
        Gaff-2.11~\cite{gaff_paper} is a general-purpose force field, ff14SB~\cite{amber14} is an established protein force field and RNA.OL3~\cite{rna_ol3} is specialized to RNA.
        The SPICE-Pubchem and SPICE-DES-Monomers datasets, containing small molecules, along with the dipeptide dataset, are subsets of the SPICE dataset~\cite{spice}. The RNA datasets feature trinucleotide states calculated with the B3LYP-D3BJ functional.
        }
    \label{tab:benchmark}
\end{table*}

\subsection{Grappa is extensible across chemical space}\label{sec:results_extensibility}

\begin{table*}[h!]
    \begin{tabular}{l c c l c  c c c c c}

\hline
\hline
\multirow{2}{*}{Dataset} & \multirow{2}{*}{Test Mols} & \multirow{2}{*}{Confs} & & \multirow{2}{*}{Grappa} & \multirow{2}{*}{Espaloma} & \multirow{2}{*}{Gaff-2.11} & \multirow{2}{*}{ff99SB-ILDN} & Mean\\
& & & &&&& & Predictor \\
\hline

\multirow{2}{*}{SPICE-Dipeptide} & \multirow{2}{*}{67} & \multirow{2}{*}{2592} & \textit{Energy} & \textbf{2.4} & 3.1 & 4.5 &  & 19.1 \\
                                 & & & \textit{{Force}}  & \textbf{5.4} & 7.8 & 12.9 &  & 38.4 \\
\hline

\multirow{2}{*}{Dipeptides-300K} & \multirow{2}{*}{72} & \multirow{2}{*}{3600} & \textit{Energy} & \textbf{2.6} &  & 4.5 & 4.1 & 7.8 \\
                                 & & & \textit{{Force}}  & \textbf{5.9} &  & 10.9 & 11.8 & 43.7 \\
\hline
\multirow{2}{*}{Dipeptides-1000K} & \multirow{2}{*}{72} & \multirow{2}{*}{2160} & \textit{Energy} & \textbf{5.4} &  & 8.6 & 8.5 & 18.9 \\
                                 & & & \textit{{Force}}  & \textbf{11.6} &  & 16.1 & 17.8 & 74.3 \\
\hline
\multirow{2}{*}{Non-Capped-Peptides} & \multirow{2}{*}{10} & \multirow{2}{*}{500} & \textit{Energy} & \textbf{2.2} &  & 4.2 & 4.0 & 7.2 \\
                                 & & & \textit{{Force}}  & \textbf{6.1} &  & 12.1 & 12.6 & 45.8 \\
\hline
\multirow{2}{*}{Radical-Dipetides} & \multirow{2}{*}{28} & \multirow{2}{*}{272} & \textit{Energy} & \textbf{3.3} &  &  &  & 8.7 \\
                                 & & & \textit{{Force}}  & \textbf{6.8} &  &  &  & 41.3 \\
\hline
\hline
\hline
\end{tabular}
    \caption{
        Accuracy of Grappa-1.3 with nonbonded parameters from ff99SB, Espaloma~0.3~\cite{espaloma0.3_paper} and established MM force fields on a subset of test molecules from the Grappa-1.3 dataset.
        The performance on small molecules and RNA is similar to the one reported in Table~\ref{tab:benchmark} and can be found in Table~\ref{tab:full_grappa}.
        As in Table~\ref{tab:benchmark}, we report the RMSE of centered energies in kcal/mol and the componentwise RMSE of forces in kcal/mol/Å.
        }
    \label{tab:grappa}
\end{table*}

If a certain kind of chemistry is not part of the training set of a machine learned force field, predictions can go awry.
While the current datasets already cover a significant part of chemical space, extensions should be straightforward and acessible.
To demonstrate Grappa's extensibility, we train it on the Grappa-1.3 dataset, which includes dipeptides sampled at 300 and 1000\,K, N- and C-terminal amino acids, dipeptides containing the non-standard residues hydroxyproline and DOPA (dihydroxyphenylalanine), and, demonstrating Grappa's capability to parametrize rather uncommon molecules, peptide radicals, which play a role in enzyme catalysis~\cite{mechanoradicals_lebrette} and as mechanoradicals~\cite{Zapp2020}.
\\

To create these additional datasets, we calculate DFT energies and gradients of states that were sampled from MD simulations performed with traditional force fields at 300\,K and 1000\,K, where we increase the temperature to 1000\,K in between the samplings of 300\,K-states to enrich conformational diversity, as described in~\ref{sec:dataset_creation}.

In Table~\ref{tab:grappa}, we report Grappa's accuracy on the Grappa-1.3 dataset.
Again, Grappa outperforms the baselines Amber99SB-ILDN and Gaff-2.11 on all peptide datasets, also for states sampled at 1000\,K, indicating that Grappa is robust under conformational perturbations that lead to out-of-equilibrium states.
For the radical peptides, there is no baseline since, to the best of our knowledge, Grappa is the first MM force field that covers this part of chemical space.
On the other datasets from Espaloma, Grappa remains competitive, as can be seen in Table~\ref{tab:full_grappa}.

The extension to the Grappa-1.3 dataset indicates that Grappa can learn new chemistries accurately without negatively affecting the performance on the original datasets.
Due to Grappa's limited complexity of input features -- only connectivity, atomic numbers and partial charges -- the chemical space accessible for parameterization is vast.
As long as a molecule has a fixed connectivity and the MM energy functional is a reasonable approximation, Grappa can be trained to predict parameters for it.

\subsection{Grappa is compatible with multiple nonbonded parameter schemes }
\label{sec:compatibility_nonbonded}

\begin{table}[h!]
    \centering
    \begin{tabular}{l c c}

\hline
\hline
Charge Model & Energy RMSE & Force RMSE \\
\hline
ff99SB & 2.55 $\pm$ 0.05 &  5.85 $\pm$ 0.06 \\
CHARMM36 & 2.58 $\pm$ 0.07 & 6.01 $\pm$ 0.07 \\
AM1-BCC & 2.79 $\pm$ 0.07  &  6.13 $\pm$ 0.07 \\
\hline
\hline
\hline
\end{tabular}
    \captionsetup{justification=raggedright,width=0.9\columnwidth}
    \caption{
        Test accuracy of Grappa-1.3 on the Dipeptides-300K dataset in kcal/mol(/Å) with sets of nonbonded parameters obtained from ff99SB, CHARMM36 and AM1-BCC charges with Lennard-Jones parameters from the OpenFF-2.2.0 force field.
        }
    \label{tab:charge_models}
\end{table}

Grappa-1.3 is trained to predict bonded parameters that are compatible with partial charges from AM1-BCC, ff99SB and CHARMM36. Table~\ref{tab:charge_models} shows that, in combination with all of these charge models, Grappa-1.3 can predict energies and forces of dipeptides to higher accuracy than established force fields (Table~\ref{tab:grappa}).
Also for partial charges from ff99SB with additional Gaussian noise of scale 0.1\,e, Grappa-1.3 is able to predict energies and forces accurately, indicating that the model is robust under small changes in the partial charges and that Grappa can thus be used in combination with charge models that are not present in the training data.
This can be partly attributed to the fact that, on the datasets considered, nonbonded interactions contribute only a small fraction to the total interaction.
For learning or evaluating the quality of nonbonded parameters by accuracy of energies and forces, we expect datasets that include intermolecular interactions to be more suitable.


\subsection{Grappa can be combined with established force fields}

\begin{table}[h!]
    \centering
    \begin{tabular}{ccc|cc}
    \hline
    \hline
    \multicolumn{3}{c|}{{Learnable contribution}} & \multicolumn{2}{c}{{RMSE [kcal/mol(/Å)]}} \\
    \hline
    Bond & Angle & Dihedral & Energy & Force \\
    \hline
\checkmark & \checkmark & \checkmark & 2.5 & 5.9 \\ 
\checkmark & \checkmark &  & 3.0 & 6.1 \\ 
\checkmark &  &  & 3.9 & 7.4 \\ 
 & \checkmark &  & 3.8 & 9.2 \\ 
 &  & \checkmark & 3.5 & 11.0 \\ 
 &  &  & 4.1 & 11.8 \\ 

    \hline
    \hline
\end{tabular}
    \captionsetup{justification=raggedright,width=0.9\columnwidth}
    \caption{
        Test accuracy of Grappa models that predict MM parameters for different interaction types. For the non-learnable types, MM parameters are taken from ff99SB-ILDN. The models are trained on the Dipeptides-300K and -1000K datasets and evaluated on Dipeptides-300K with ff99SB partial charges.
        }
        \label{tab:amber_ablation}
\end{table}

Grappa's modular formulation allows for learning only a certain type of interaction while keeping the others fixed to those of an established force field.
To quantify the improvement of Grappa over the widely used protein force field Amber ff99SB-ILDN~\cite{amber99sbildn_paper}, we train Grappa models that predict only certain types of MM parameters and evaluate their accuracy in Table~\ref{tab:amber_ablation}.
While for each contribution, replacing the ff99DB-ILDN parameters with those predicted by Grappa improves the accuracy, the bond and angle contributions have the most potential for improvement.
\todo{for final version: how can one combine grappa parameters with amber/charmm bonded?}

\subsection{Grappa is interpretable}\label{sec:interpretable}

\paragraph{MM parameters}
In the classical mechanics potential functions used in MM, a physical meaning can be attributed to parameters. Hence, badly assigned parameters could be noticeable, e.g. if a bond is much shorter or stiffer than bonds between similar atoms or the amplitude of a proper dihedral is untypically high. 
Apart from exposing the parameters in the simulation files, Grappa also provides figures of the parameter distributions and, if available, comparisons to parameters from an established force field upon parametrization.
For the case of the protein ubiquitin considered in section~\ref{sec:simulation_stability}, parameter distributions and a comparison to ff99-SBILDN are depicted in Figure~\ref{fig:parameter_comparison}.
While the established protein force field and Grappa-1.3 predict similar bond distances, we observe deviations for other types of parameters.

\todo{violin plot (and absolute comparison) ubq, combined with latent space in large fig}

\paragraph{Latent space}
To generalize across the broad range of chemical space covered by the training set, Grappa has to learn highly informative atom embeddings that capture the local environment in the molecular graph.
For interpreting this latent representation, we apply a two-dimensional principal component analysis on a set of predicted atom embeddings (\ref{eq:atom:embed}) of test molecules (Figure~\ref{fig:scatter_and_pca}b).
It turns out that the two principal components can be related to main group and period of the periodic table of elements, whose structure is learned implicitly by Grappa.

\section{Force field validation}\label{sec:validation}

Especially for large biomolecules, force fields are not only required to predict energies and forces as accurately as possible, but their predictions should behave in such a way that certain macroscopic properties such as stability and consistency with experimentally determined folding states are fulfilled in MD simulations.
Grappa is well suited to overcome the gap between empirically validated, established protein force fields and machine learned force fields.

In this section, we demonstrate that Grappa is on par with established force fields when it comes to MD simulations of proteins over timescales of hundreds of nanoseconds and that it captures the physics that cause stability of protein folds.
All simulations mentioned in this section were performed with GROMACS, for which the Grappa package\footnote{
\url{https://github.com/graeter-group/grappa}
} provides a command line interface, briefly described in~\ref{sec:package}.
Using this interface (or a similar one for OpenMM), large biomolecular systems with over 100,000 atoms are parameterized within minutes on a CPU (Figure~\ref{fig:runtime}).

\subsection{Grappa keeps large proteins stable during MD}
A key quality check for a protein force field is if a native protein structure remains stable over time during an MD simulation.
QM methods and E(3) equivariant neural networks suffer from the problem that while their end-to-end predictions are accurate, 
they might fail in yielding stable simulations because of diverging gradients of the potential energy surface when extrapolating beyond regions covered by the training dataset.
Particularly over long timescales, the system might drift out of the range of validity of the 
model and small errors might accumulate, leading to instabilities of the system and a potential crash of the simulation.
MM with its interpretable, physics-inspired energy function, with parameters which are conformation-independent, ensures a higher degree of stability, and we indeed did not observe simulation crashes. 

\begin{figure}[h]
    \centering
    \includegraphics[width=\columnwidth]{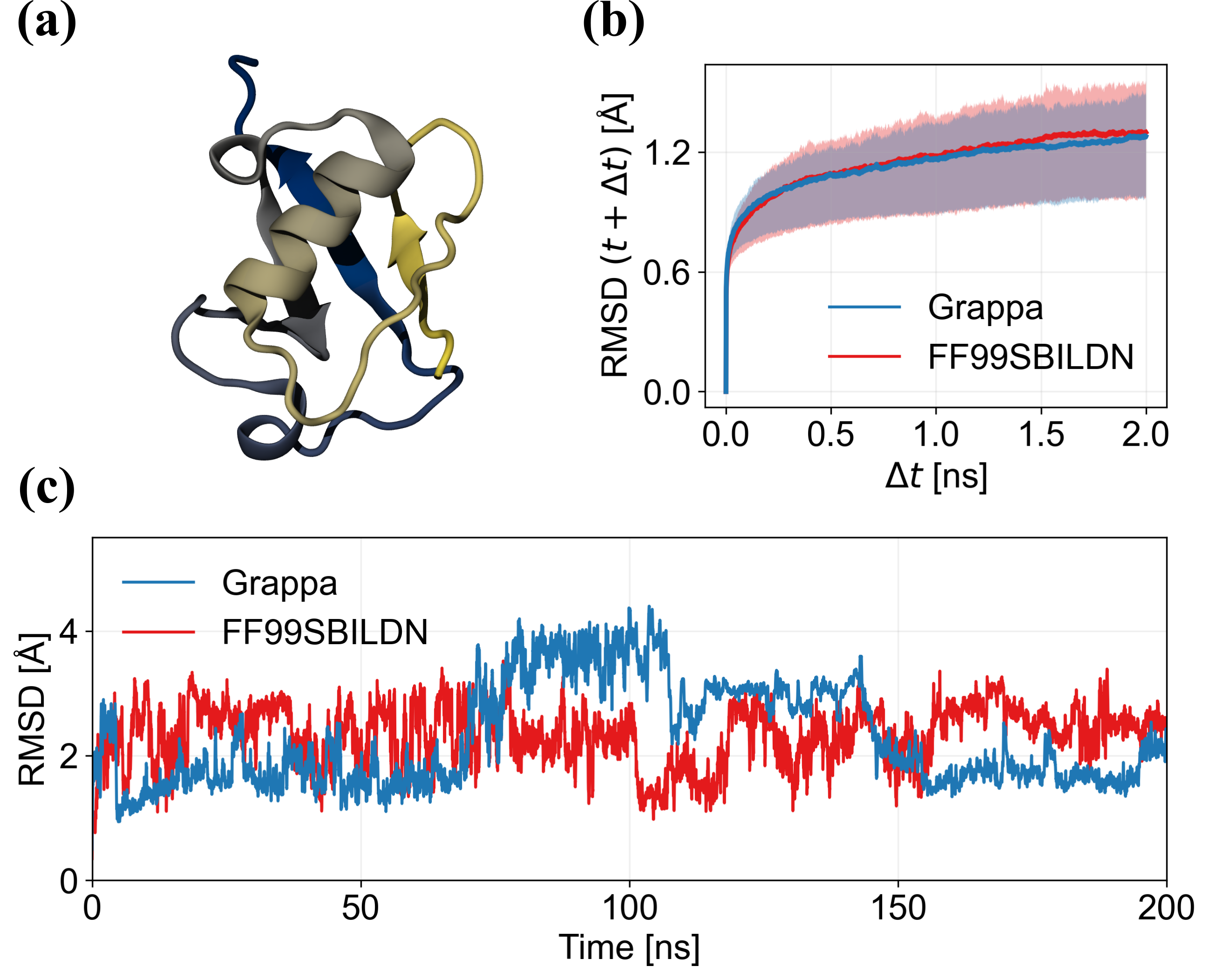}
    \caption{
    \textbf{(a)} The protein ubiquitin with color-coded sequence position.
    \textbf{(c)} The mean C-alpha root mean square deviation (RMSD) and its 25th and 75th percentile of 1000 random pairs of frames that are separated by the time difference $\Delta t$.
    \textbf{(b)} C-alpha RMSD from the initial state during MD simulation of ubiquitin in water with Grappa.
    }
    \label{fig:stability_rmsd}
\end{figure}

However, since it is not ensured a priori that macrostates such as protein folds remain stable during MD, we next assess Grappa's capability of keeping protein folds close to their experimentally determined structure.
As example system, we use ubiquitin (PDB ID: 1UBQ), which is a protein with 76 amino acids whose fold contains a beta-sheet and an alpha-helix.
We perform a 200 nanoseconds MD simulation of ubiquitin in aqueous solution with Grappa and Amber ff99SB-ILDN~\cite{amber99sbildn_paper}, for which details are given in~\ref{sec:simulation_stability}.
We find that the C-alpha RMSD from the initial state is bounded by about 4\,Å in 200\,ns simulation with both Grappa and Amber ff99SB-ILDN (Figure~\ref{fig:stability_rmsd}c), indicating that the folded state of the protein is stable.
Structural fluctuations on timescales of up to 2\,ns are of similar magnitude (Figure~\ref{fig:stability_rmsd}b).

\subsection{Grappa is orders of magnitude more resource-efficient than E3 equivariant models}
\label{sec:efficiency}

Efficiency of force fields is crucial; otherwise the computational cost for simulating large systems on long timescales can prohibit their application in simulations of many systems of interest.
To demonstrate Grappa's efficiency, which it inherits from molecular mechanics, and to showcase its capability of jointly parametrizing RNA and proteins, we simulate the virus STMV~\cite{Amber23Benchmark} in solution -- a system with approximately one million atoms, visualized in Fig.~\ref{fig:virus}~(a).
\\

A recently proposed machine learned force field relying on an E(3) equivariant neural network with state-of-the-art accuracy and efficiency, Allegro~\cite{allegro}, has been shown to be capable of performing near-quantum-accuracy simulations of systems of unprecedented size.
This is achieved by using a strictly local architecture, which allows parallelization across thousands of GPUs.
Since they are not constrained by the energy functional of MM, E3 equivariant models like Allegro, NequIP~\cite{nequip} or MACE~\cite{mace} have greatly improved accuracy but are orders of magnitude more expensive than MM force fields like Grappa.

For the system at hand, Allegro achieves a performance of 106 timesteps per second on 4,000 A100 GPUs~\cite{stmv-allegro}. 
For Grappa, at reduced accuracy and incapable of describing topological changes directly, we measure a performance of 101 timesteps per second on a single A100 GPU in GROMACS~\cite{gromacs} without system-specific optimizations.

\section{Conclusions}

With Grappa, we propose a machine learning framework for molecular mechanics force fields with state-of-the-art accuracy that can be used seamlessly in established MD engines.
Our experiments show that Grappa is transferable to large biomolecules like proteins and viruses, which sets the stage for large scale simulations at improved accuracy, with the same computational efficiency as established MM force fields.

Unlike most traditional MM force fields, Grappa is not specialized to a certain chemical species, but offers consistent parameterization of molecules with different chemistries, such as a wide range of small molecules, peptides and nucleotides, with a single model at a higher level of accuracy than previous MM force fields.
Grappa not only enhances existing parameter sets, but also reaches previously inaccesible regions of chemical space, such as non proteinogenic amino acids or protein radicals.

Since current state-of-the-art E(3) equivariant neural network potentials like Allegro are several orders of magnitude more expensive than molecular mechanics, we regard machine learned molecular mechanics approaches like Grappa as suitable method for simulating large systems, especially if computational ressources are limited or when it is not necessary to simulate all parts of a system at quantum accuracy.

\section{Outlook}
While Grappa achieves high computational efficiency through molecular mechanics, it also inherits its fundamental limitations, including restricted accuracy and the inability to directly describe chemical reactions.
However, Grappa is well suited for efficient reparametrization of large molecules that undergo local topological changes induced by chemical reactions, e.g. in kinetic Monte Carlo simulations~\cite{kimmdy}. This is due to both, Grappa's finite field-of-view, which ensures locality on the molecular graph, as well as the lack of hand-crafted chemical features, which allows to apply the parametrization to cut-out regions of molecules. 

Grappa is a general and versatile framework to obtain MM parameters for a broad range of molecules and can be seamlessly extended to new chemistries.
To this end, the Grappa package implements workflows for retraining on the provided and custom new datasets.

While we consider the consistency with established nonbonded parameters an advantage of Grappa, an extension of the framework to nonbonded parameters is a straightforward next step. It would render Grappa fully independent from traditional MM force fields and could potentially improve its accuracy further.
\section{Reproducibility statement}
Grappa is released as open source software under the GNU General Public License v3.0
\footnote{\url{https://github.com/graeter-group/grappa}.
}
along with the Grappa dataset and SMILES strings used for the train-val-test partition and configuration files to reproduce Table~\ref{tab:benchmark} and Table~\ref{tab:grappa}.

Repositories for reproducing the MD simulations
\footnote{
\url{https://github.com/LeifSeute/validate-grappa}
}
and creating the Grappa-1.3 datasets
\footnote{
\url{https://github.com/LeifSeute/grappa-data-creation}
}
are publicly available.

\section*{Author Contributions}

\textbf{Leif Seute:} Methodology, Conceptualization, Software, Writing - Original Draft, Investigation, Visualization, Data Curation\\
\textbf{Eric Hartmann:} Conceptualization, Software, Verification, Visualization, Writing - Review and editing \\
\textbf{Jan St\"uhmer:} Formal analysis, Writing - Review and editing \\
\textbf{Frauke Gr\"ater:} Supervision, Conceptualization, Writing - Review and editing \\


\section*{Acknowledgements}


This research was supported by the Klaus Tschira Foundation.
The project has also received funding from the European Research Council (ERC) under the European Union’s Horizon 2020 research and innovation programme (grant agreement No. 101002812) [F.G.].
We thank Kai Riedmiller and Jannik Buhr for application and software related discussions, Fabian Gr\"unewald, Camilo Aponte-Santamaria and Vsevolod Viliuga for discussions on future applications and Daniel Sucerquia for conversations on the role of internal coordinates in MM.




\bibliography{libfile}
\bibliographystyle{icml_style/icml2024}

\appendix

\renewcommand\thefigure{A\arabic{figure}}
\renewcommand\thetable{A\arabic{table}}

\section{Appendix}

\subsection{The Grappa package}\label{sec:package}
Grappa utilizes PyTorch~\cite{pytorch} and DGL~\cite{dgl} in its implementation.
The package is released as open source software under the GNU General Public License v3.0 and is available on GitHub
\footnote{
\url{https://github.com/graeter-group/grappa}
}
and on PyPi:
\begin{small}
\begin{verbatim}
pip install grappa-ff
\end{verbatim}
\end{small}

It brings a command line interface to reparametrize a GROMACS topology file obtained from a traditional force field:
\begin{small}
\begin{verbatim}
grappa_gmx -f topol.top -o new_topol.top
\end{verbatim}
\end{small}

For OpenMM, we provide a class that can be used to reparametrize a system with Grappa:
\begin{python}
from grappa import OpenmmGrappa

# use a traditional OpenMM forcefield
# to obtain a system from your topology
top, system = ...

# download a pretrained grappa model
ff = OpenmmGrappa.from_tag('grappa-1.3')

# re-parametrize the system using grappa
system = ff.parametrize_system(system, top)

# continue with usual workflow
\end{python}

Pretrained released models can be accessed by using a tag; the model weights are downloaded automatically from the respective release on GitHub.
Released models also contain a dictionary with results on test datasets, a list of identifiers for molecules used for training and validation and a configuration file to reproduce training on the same dataset.
Further details on the package can be found on GitHub.

\begin{figure}[h!]
    \centering
    \includegraphics[width=0.9\columnwidth]{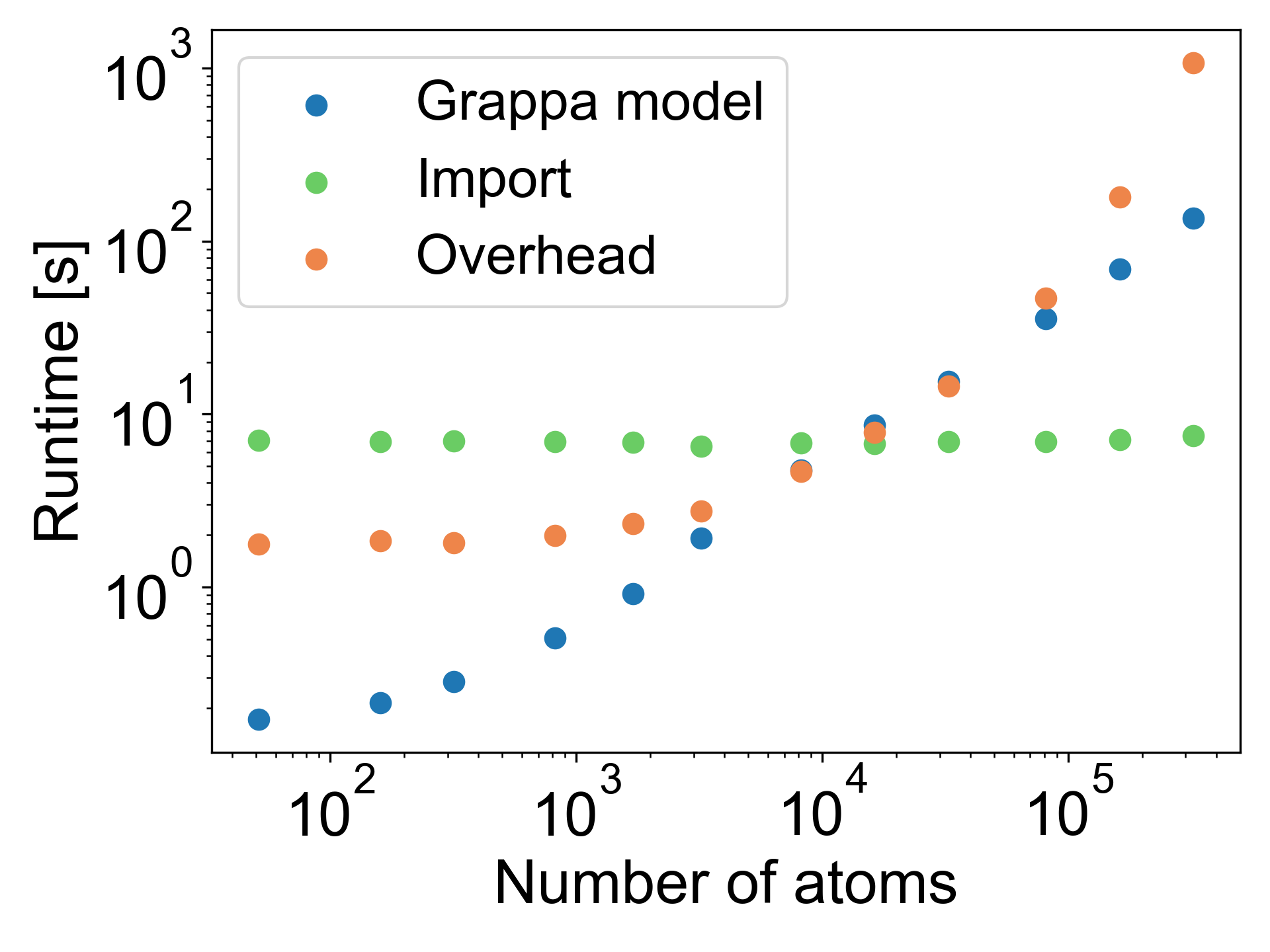}
    \caption{
    For molecules with up to 300,000 atoms, the runtime of a parameterization with the Grappa package in CPU mode is largely due to overhead.
    Due to its finite field-of-view, parametrizations can also be parallelised across several nodes by splitting the molecular graph into subgraphs if necessary.
    }
    \label{fig:runtime}
\end{figure}

\subsection{Model architecture}\label{sec:appendix_model_architecture}

In this section, we provide a detailed description of the architecture of Grappa, including the graph attentional neural network and the symmetric transformer.

\subsubsection{Graph attentional neural network}
\label{sec:gnn}

Grappa's graph attentional neural network (Figure~\ref{fig:gnn}) is a modification of the transformer architecture~\cite{transformer} for graphs, similar to the Graph Transformer Network~\cite{graph_transformer}, but e.g. omits the locality-breaking graph Laplacian eigenvectors as positional encoding.
Instead, we encode the graph structure by constraining the attention mechanism to edges of the graph as in GAT~\cite{gat_original}, which enforces locality: In each attention update only neighboring nodes can influence each other.
We use a multi-head dot-product attention mechanism and a 2-layer MLP as feed-forward network, where the hidden feature dimension is four times the node feature dimension.

After initializing the node features $\nu_i$ as described below, we apply a single nodewise linear layer followed by an exponential linear unit (ELU)~\cite{elu},
\begin{align}
    \nu_{i} &\leftarrow \text{ELU}\left( W \nu_{i} + b \right), \: W \in \mathbb{R}^{d \times d_\text{in}}, \: b \in \mathbb{R}^{d},
\end{align}
Then, we apply $L_{\text{GNN}}$ graph attentional layers, each of which is given by
\begin{align}
     \nu_i &\leftarrow \text{LayerNorm}(\nu_i), \\
     e_{ii'}^{(k)} &= \left( W^{(k)} \nu_i \right) \cdot \left( W^{(k)} \nu_{i'} \right), \\
     \text{att}^{(k)}_{ii'} &= \frac{\exp\left(e_{ii'}^{(k)}\right)}{\sum_{n \in \mathcal{N}(i)} \exp\left(e_{in}^{(k)}\right)}, \\
     V^{(k)}_i &= \sum_{i' \in \mathcal{N}(i)} \text{att}^{(k)}_{ii'} W^{(k)} \nu_{i'}, \\
     V_i &\leftarrow \left( W_O \text{Concat}\left[ V^{(1)}_i, \ldots, V^{(N_{\text{heads}})}_i \right] \right), \\
     \nu_i &\leftarrow \text{LayerNorm}(V_i) + \nu_i, \label{eq:gnn:-2}\\
     \nu_i &\leftarrow \text{ELU}\left( W_2 \: \text{ELU}\left( W_1 \nu_i + b_1 \right) + b_2 \right) + \nu_i, \label{eq:gnn:-1}
\end{align}
where $\mathcal{N}(i)$ is the set of neighbors of node $i$ including the node itself, $(k)$ denotes the attention head and where the weights
 \begin{align}
     W^{(k)} &\in \mathbb{R}^{d \times d/N_{\text{heads}}}, \\
     W_O &\in \mathbb{R}^{d \times d}, \\
     W_1 &\in \mathbb{R}^{4d \times d}, \\
     W_2 &\in \mathbb{R}^{d \times 4d}.
\end{align}
and biases $b_1, b_2$ are learnable and independent for each layer.
We use dropout~\cite{dropout} on the node feature update in~\eqref{eq:gnn:-2} and ~\eqref{eq:gnn:-1}.
Finally, we project onto the output dimension $d_\text{emb}$ by a linear layer,
\begin{align}
    \nu_i &\leftarrow W \nu_i, W \in \mathbb{R}^{d_\text{emb} \times d}.
\end{align}

\begin{figure}
     \centering
     \resizebox{0.99\columnwidth}{!}{
          \def\nodedistance{0.8}
\def\halfnodedistance{0.4}

\def\xshiftgraphs{2.2}
\def\yshiftgraphin{6.2}
\def\yshiftgraphout{-8.7}
\def\colsep{1.3}

\begin{tikzpicture}[
    node distance=\nodedistance cm and 1.2cm,
    very thick,
    process/.style={
        rectangle,
        rounded corners,
        minimum width=3cm,
        minimum height=1cm,
        text centered,
        draw=black,
        top color=mycolor!30,
        bottom color=mycolor!35,
        blur shadow={shadow blur steps=3}
    },
    ln/.style={
        rectangle,
        rounded corners,
        minimum width=3cm,
        minimum height=1cm,
        text centered,
        draw=black,
        top color=mycolor!20,
        bottom color=mycolor!25,
        blur shadow={shadow blur steps=3}
    },
    other_style/.style={
        rectangle, 
        rounded corners, 
        minimum width=3cm, 
        minimum height=1cm, 
        text centered, 
        draw=black, 
        top color=black!7, 
        bottom color=black!10,
        blur shadow={shadow blur steps=3}
    },
    arrow/.style={
        ->,
        >=stealth,
        very thick,
        rounded corners
    },
    arrow_nohead/.style={
        -,
        very thick,
        rounded corners
    },
    plus/.style={
        circle,
        fill=white,
        draw,
        inner sep=1pt
    },
    smallnode/.style={
        circle,
        draw,
        minimum size=0.4cm,
        fill=grey!30
    },
    smallnode_highlight/.style={
        smallnode,
        fill=red
    },
    edge/.style={
        very thick,
        rounded corners
    },
    linear_style/.style={
        process, fill=blue!30
    },
    attention_style/.style={
        process, fill=blue!50
    },
    sum_style/.style={
        process, fill=orange!40
    },
    concat_style/.style={
        process, fill=orange!20
    }
]

\tikzstyle{edge} = [very thick,-, rounded corners]
\tikzstyle{plus} = [circle, fill=white, draw, inner sep=1pt]

\moleculargraph{-\xshiftgraphs}{\yshiftgraphin}

\matrix (m) [column sep=\colsep cm, row sep=\nodedistance cm, align=center] {
    \node [other_style] (g) {Molecular Graph}; & \node [other_style] (atomic_numbers) {Atomic Numbers};\\
    & \node [process] (lin1) {Linear}; \\
    & \node [ln] (ln1) {Layer Norm}; \\
    & \node [process, minimum height=1.2cm] (att) {\textbf{Multihead}\\ \textbf{Graph Attention}}; \\
    & \node [ln] (ln2) {Layer Norm}; \\
    & \node [process, minimum height=1.2cm] (ff) {\textbf{Nodewise}\\ \textbf{Feed Forward}}; \\
};

\node [other_style, below=of ff, yshift=-0.5cm] (atom_embeddings) {Atom Embeddings};

\draw [arrow] (atomic_numbers) -- (lin1);
\draw [arrow] (lin1) -- (ln1);
\draw [arrow] (ln1) -- (att);
\draw [edge] (att) -- (ln2);
\draw [arrow] (ln2) -- (ff);
\draw [arrow] (ff) -- (atom_embeddings);

\draw [edge] (g) |- (att);

\draw [arrow] (g.south) |- ++(0,-\halfnodedistance) -| (lin1.north);    

\draw [edge] ([yshift=\halfnodedistance cm]att.north) -- ++(2.3,0) |- ([yshift=-\halfnodedistance cm]att.south);
\draw [edge] ([yshift=\halfnodedistance cm]ff.north) -- ++(2.3,0) |- ([yshift=-\halfnodedistance cm]ff.south);

\node [plus] at ([yshift=-\halfnodedistance cm]att.south) {+};
\node [plus] at ([yshift=-\halfnodedistance cm]ff.south) {+};

\begin{scope}[on background layer]
    \draw[rounded corners, thick, draw=black, fill=black!10] ([shift={(-1.3cm,\halfnodedistance cm)}]ln1.north west) rectangle ([shift={(1.3cm,-\nodedistance cm)}]ff.south east) coordinate (rect_se) at (current path bounding box.south east);
\end{scope}

\node [left=0.5cm of ln1, anchor=west, scale=1.5, yshift=0.35cm, xshift=-0.55cm] {\(n \times \)};

\atomembeddinggraph{\xshiftgraphs}{\yshiftgraphout}

\end{tikzpicture}
     }
     \caption{
         Grappa's graph attentional neural network as described in Section~\ref{sec:gnn}.
         }
     \label{fig:gnn}
\end{figure}

As initial node features we choose a one-hot encoding of the atomic number, a one-hot encoding of whether the node is in a loop, a one-hot encoding of the potential loop size and the degree of the node.
Since Grappa only predicts bonded parameters, we have to ensure consistency with the nonbonded parameters from the traditional force field, which is why we one-hot encode the traditional force field used for the nonbonded contribution of the respective state if we train on a dataset that contains data from different nonbonded methods.
We also include the partial charges (the scalar value concatenated with a 16-dimensional binning between -2$e$ and 2$e$) as node input features, which allows us to describe differently charged conformations of the same molecule without resorting to global or graph-symmetry breaking features like the formal charge.

\subsubsection{Symmetric transformer}\label{sec:symmetric_transformer}

The graph attentional neural network is followed by four (parallel) symmetric transformers (Figure~\ref{fig:symmetric_transformer}), one for each type of interaction, that is one for bond, one for angle, one for torsions and one for improper dihedral parameters.
For each interaction (e.g. for each angle in the molecular graph), the node embeddings of the atoms involved is mapped to a set of respective MM parameters (e.g. equilibrium angle and force constant).

As described in Section~\ref{sec:grappa}, we add a symmetric positional encoding to the node features, as in (\ref{eq:pos_enc}), that is invariant under the desired set of permutations but can break symmetries that are not necessary to make the model more expressive while keeping equivariance under the permutations that we use for invariant pooling (\ref{eq:pooling}) later on.
From the symmetries Eqs.~\ref{eq:bond_symm}-\ref{eq:torsion_symm}, and from the considerations for improper dihedrals in Section~\ref{sec:impropers}, we can derive the following positional encodings
\begin{align}
    \text{PE}_{\text{angle}} &= (0,1,0),\\
    \text{PE}_{\text{torsion}} &= (0,1,1,0),\\
    \text{PE}_{\text{improper}} &= (0,1,1,0).
\end{align}
The bond positional encoding is not needed because the permutations of the bond parameters are the full group $S_2$ and there is no symmetry that needs to be broken.
For torsions and impropers, the positional encoding does not break all unnecessary symmetries, for example the symmetry $ijkl \rightarrow ljki$ is still present with positional encoding, and will be broken later.

We then apply a transformer acting one nodes that represent the atoms involved in the interaction, that is, after a projection on features of dimension $d_T$,
\begin{align}
    \nu_{i} &\leftarrow \text{ELU}\left( W \nu_{i} + b \right), \: W \in \mathbb{R}^{d_T \times d_\text{emb}}, \: b \in \mathbb{R}^{d_T},
\end{align}
we apply $L_{\text{T}}$ transformer layers, each of which is given by scaled dot-product attention with $N_{\text{T-heads}}$ heads and a 2-layer MLP with hidden dimension $4d_T$, skip connections, dropout and layer normalization as in the original transformer architecture~\cite{transformer}.

We then apply a symmetry pooling operation by passing permuted versions of concatenated node embeddings to a $L_{\text{pool}}$-layer MLP with hidden dimension $d_{\text{pool}}$ and summing over all permutations $\sigma$ in the respective symmetry group $\mathcal{P}$,
\begin{align}
    z_{ij\dots} = \sum_{\sigma\in \mathcal{P}} \text{MLP}\left(\left[\nu_{\sigma(i)}, \nu_{\sigma(j)}, \dots\right]\right), \label{eq:app_pooling}
\end{align}
The two dimensional scores $z_{ij}$ for bonds and $z_{ijk}$ for angles are finally mapped to the range of the respective set of MM parameters by scaled and shifted versions of ELU and the sigmoid function,
\begin{align}
    k_{ij} &= \text{ToPos}\left(z_{ij,\,0}\right), \\
    r_{ij}^{(0)} &= \text{ToPos}\left(z_{ij,\,1}\right), \\
    k_{ijk} &= \text{ToPos}\left(z_{ijk,\,0}\right), \\
    \theta_{ijk}^{(0)} &= \text{ToRange}_{(0,\pi)}\left(z_{ijk,\,0}\right),
\end{align}
where $\text{ToPos}$ and $\text{ToRange}_{(0,\pi)}$ are defined in Section~\ref{sec:scaling} and $z_{\dots,\,n}$ denotes the $n$-th entry of the score vector $z_{\dots}$.
For dihedral force constants, the $2n_\text{periodicity}$ dimensional scores $z_{ijkl}$ are fed through a sigmoid gate to make the model more expressive for small force constants,
\begin{align}
    k_{ijkl,\,m} &= z_{ijkl,\,2m+1} \times\text{sigmoid}\left(z_{ijkl,\,2m}\right).
\end{align}

\subsubsection{Scaling of neural network outputs}\label{sec:scaling}

To map the scores predicted from the model to the range of the respective MM parameter, we use  scaled and shifted versions of the sigmoid function for mapping to a specified interval and of ELU for mapping to positive values.
We choose these functions because they are differentiable and, in contrast to the exponential, only grow linearly for large inputs, which can help to stabilize training.
To allow the model to predict scores that are of order unity, or approximately normally distributed, we choose the scaling and shifting parameters in such a way that a normally distributed input would lead to an output distribution whose mean and standard deviation are close to a given target mean $\mu$ and standard deviation $\sigma$.
We choose this target standard deviation by calculating mean and standard deviation of traditional MM force field parameters on a given subset of the training data, except for equilibrium angles, where we restrict $\mu$ to $\pi/2$.

For mapping to positive values, we use
\begin{align}
    \text{ToPos}[\mu,\sigma] \left(z\right) 
    =
    \sigma \left(\text{ELU}\left(\frac{\mu}{\sigma} + x - 1\right) + 1\right),
\end{align}
and for mapping to a specified interval $(0,\gamma)$, we restrict ourselves to the case $\mu=\gamma/2$ use
\begin{align}
    \text{ToRange}[\gamma/2,\sigma]_{(0,\gamma)} \left(z\right)
    =
    \gamma \: \text{sigmoid}\left(\frac{4\sigma}{\gamma}x\right).
\end{align}
We can see that the mean and standard deviation of the output are indeed close to the target mean and standard deviation if we consider the asymptotic behavior of the sigmoid function and ELU as their input approaches zero,
\begin{align}
    \text{ELU}(x) & \sim x + \mathcal{O}\left(x^2\right) \quad \text{as} \quad x \rightarrow 0, \\
    \text{sigmoid}(x) & \sim \frac{1}{2} + \frac{x}{4} + \mathcal{O}\left(x^2\right) \quad \text{as} \quad x \rightarrow 0,
\end{align}
which follows from the Taylor expansion of the ELU and sigmoid functions around zero.
Thus, using that $z$ has vanishing mean and unit standard deviation, we have indeed, to first order in $z$,
\begin{align}
    <\text{ToRange}\left(z\right)> &\approx \gamma \left(1/2 + {4\sigma}/{\gamma}<z>\right)\notag =
    {\gamma}/{2} \notag
\end{align}
and
\begin{align}
    \text{Var}\left[
        \text{ToRange} \left(z\right)
    \right]
    \approx <\gamma^2 {\left(1/2 + \frac{\sigma}{\gamma}z\right)}^2> - \gamma^2/4 \notag\\
    = \gamma^2\left(1/4 + \frac{\sigma}{\gamma}<z> + {\sigma^2}/{\gamma^2}<{z}^2> - 1/4\right) = \sigma^2.\notag
\end{align}

For mapping to positive values, we expand ELU in $\mu/\sigma - 1 + z$ around zero, which is reasonable if $\mu/\sigma \approx 1$, and find in analogy that target mean and standard deviation are recovered to first order in $\mu/\sigma - 1 + z$.

\subsection{Hyperparameters and training details}\label{sec:hyperparams}

\begin{table}[h!]
    \centering
    \begin{tabular}{llc}
        \toprule
        Hyperparameter & Variable & Value \\
        \midrule
        \multicolumn{3}{c}{} \\
        \multicolumn{3}{c}{Graph Neural Network} \\
        \midrule
        Layers & $L_{\text{GNN}}$ & 7\\
        Hidden dimension & $d$ & 512\\
        Attention heads & $N_{\text{heads}}$ & 16\\
        Embedding dimension & $d_{\text{emb}}$ & 256\\
        Dropout probability &  & 0.3\\
        \multicolumn{3}{c}{} \\
        \multicolumn{3}{c}{Symmetric Transformer} \\
        \midrule
        Transformer layers & $L_T$ & 3\\
        Transformer hidden dim & $d_T$ & 512\\
        Attention heads & $N_{\text{T-heads}}$ & 8\\
        Pooling layers & $L_{\text{pool}}$ & 3\\
        Pooling hidden dim & $d_{\text{pool}}$ & 256\\
        Dropout probability &  & 0.5\\
        \multicolumn{3}{c}{} \\
        \multicolumn{3}{c}{Training setup} \\
        \midrule
        Learning rate & & $1.5\times 10^{-5}$\\
        Molecules per batch & & 32\\
        States per molecule & & 32\\
        Force weight & $\lambda_{F}$ & 0.8\\
        Dihedral $L_2$ weight & $\lambda_{\text{dih}}$ & $10^{-3}$\\
        Traditional MM weight & $\lambda_{\text{trad}}$ & 0\\
    \end{tabular}
    \caption{
        The hyperparameters used for training the Grappa models in this work.
        }
    \label{tab:hyperparams}
\end{table}

For all models discussed in this work, we use the hyperparameters listed in Table~\ref{tab:hyperparams} and a dihedral periodicity of $n_{\text{periodicity}}=3$.
All relative weights between energies, forces and MM parameters rely on units formed by kcal/mol, Å and radian.
We train the models using the Adam~\cite{adam} optimizer with $\epsilon=10^{-8}$, $\beta_1=0.9$, $\beta_2=0.999$.

First, we train for 2 epochs on traditional MM parameters only, which can be interpreted as a form of pretraining that is efficient in bringing the model in a state where the gradients of the QM-part of the loss functions are informative.
We start the training on energies and forces with $\lambda_{\text{trad}}=10^{-3}$, which we set to zero after 100 epochs, preventing the model from converging towards unphysical local minima of the QM-part of the loss function.
When changing weights of terms of the loss function as described above, we re-initialize the optimizer with random moments and apply a warm restart~\cite{warm_restart}, during which we linearly increase the learning rate in 500 training steps to decorrelate the optimizer state from the gradient updates of the previous loss function.
Besides pretraining on traditional MM parameters directly, we have found penalizing large torsion and improper force constants $\xi^{(\text{dih})}$ to benefit generalizability across conformational space.

As validation loss, we use a linear combination of the energy and force root mean squared error (RMSE) on the sub-datasets (the rows in Table~\ref{tab:benchmark}), averaged over all sub-datasets, with a weight of 1 for the energy RMSE and 3 for the force RMSE to prevent overfitting on a molecule type.
We use this validation loss for early stopping and for learning rate scheduling.
For the model trained on the Espaloma dataset (Table~\ref{tab:benchmark}), the validation loss did not improve after 1000 epochs, which corresponds to about 25 hours of training on an A100 GPU.

\subsection{Improper dihedrals}\label{sec:impropers}

\begin{figure}
    \centering

    \tikzset{smallnode_style/.style={
        circle,
        minimum size=0.2cm,
        text centered, 
        draw, 
        top color=mycolor!30,
        bottom color=mycolor!35,
        blur shadow={shadow blur steps=3}
    }}

    \begin{subfigure}{0.45\columnwidth}
        \centering
        \begin{tikzpicture}
            \node[smallnode_style] (i1) at (0,0) {i};
            \node[smallnode_style] (j1) at (1.1,0) {j};
            \draw (i1) -- (j1);
        \end{tikzpicture}
        \caption{Bond}
    \end{subfigure}%
    \hfill
    \begin{subfigure}{0.45\columnwidth}
        \centering
        \begin{tikzpicture}
            \node[smallnode_style] (i2) at (0,-0.5) {i};
            \node[smallnode_style] (j2) at (1,0) {j};
            \node[smallnode_style] (k2) at (2,-0.5) {k};
            \draw (i2) -- (j2) -- (k2);
        \end{tikzpicture}
        \caption{Angle}
    \end{subfigure}

    \vspace{5mm}

    \begin{subfigure}{0.45\columnwidth}
      \centering
      \begin{tikzpicture}
        \node[smallnode_style] (i3) at (0,0.5) {i};
        \node[smallnode_style] (j3) at (1,-0.0) {j};
        \node[smallnode_style] (k3) at (2,0.5) {k};
        \node[smallnode_style] (l3) at (3,-0.0) {l};
        
        \draw (i3) -- (j3) -- (k3) -- (l3);
    
        \node (invisible1) at (0,-.5) {};
        
      \end{tikzpicture}
      \caption{Torsion}
    \end{subfigure}%
    \hfill
    \begin{subfigure}{0.45\columnwidth}
        \centering
        \begin{tikzpicture}
            \node[smallnode_style] (i4) at (0,0.5) {i};
            \node[smallnode_style] (j4) at (2,0.5) {l};
            \node[smallnode_style] (k4) at (1,0) {k};
            \node[smallnode_style] (l4) at (1,-1) {j};
            \draw (i4) -- (k4) -- (j4);
            \draw (k4) -- (l4);
        \end{tikzpicture}
        \caption{Improper}
    \end{subfigure}
      
    \caption{The subgraphs of the molecular graph that correspond to bonded MM interactions}
    \label{fig:subgraphs}
\end{figure}

As discussed in Section~\ref{sec:grappa}, we postulate that the energy contributions from the different interactions are symmetric under certain permutations of the embeddings and spatial positions of the atoms involved.
These permutations are given by the isomorphisms of the respective subgraph that describes the interaction.
In the case of angles, bonds and torsions, these symmetries are given by Eqs.~\ref{eq:bond_symm}-\ref{eq:torsion_symm}, for improper dihedrals, the symmetries are given by all six permutations that leave the central atom invariant, as can be seen from the structure of the subgraphs in Figure~\ref{fig:subgraphs}.
For bonds, angles and torsions, the interaction coordinates distance, angle and the cosine of the dihedral angle are invariant under the respective permutations, which is why we can construct an invariant energy contribution simply by enforcing invariance of the MM parameters under the respective permutations.

For improper dihedrals, this is not the case, as the dihedral angle is not invariant under all six permutations mentioned above and a model with improper force constants that are invariant under these permutations would not have invariant improper energy contributions.
In Grappa, we solve this problem by introducing more terms for improper dihedrals. For example
\begin{equation}
    E_{\text{improper}} = k_{ijkl}\cos(\phi_{ijkl}) + k_{jikl}\cos(\phi_{jikl}),
\end{equation}
is invariant under the permutation $ijkl \rightarrow jikl$, which can be generalized to all six improper permutations.
It turns out that we can reduce the number of additional terms to two by using another symmetry of the dihedral angle,
\begin{equation}
    \cos\left(\phi_{ijkl}\right) = \cos\left(\phi_{ljki}\right),
\end{equation}
which can be seen from the formula for the dihedral angle in~\cite{dihedral_paper}.
This symmetry allows us to identify pairs of terms that are transformed into each other under the permutation $ijkl \rightarrow ljki$ and use a force constant that is invariant under this permutation,
\begin{equation}
    k_{ijkl}^{(\text{improper})} = k_{ljki}^{(\text{improper})}. \label{eq:improper_symm}
\end{equation}

\subsection{Datasets}\label{sec:dataset_creation}

\subsubsection{Espaloma dataset}
We reproduce the dataset used to train and evaluate Espaloma~0.3~\cite{espaloma0.3_paper} by downloading the published, preprocessed data\footnote{
\url{https://github.com/graeter-group/grappa/blob/master/dataset_creation/get_espaloma_split/load_esp_ds.py}
}.
The preprocessing entails filtering out high-energy conformations as detailed in \cite{espaloma0.3_paper}.
As in Espaloma, our train-val-test partitioning procedure relies on the unique isomeric SMILES strings of the molecules, which are included in the dataset.
We reproduce\footnote{
\url{https://github.com/graeter-group/grappa/blob/master/dataset_creation/get_espaloma_split/load_esp_ds.py}
}
the partitioning from Espaloma~0.3\footnote{
\url{https://github.com/choderalab/refit-espaloma/blob/main/openff-default/02-train/joint-improper-charge/charge-weight-1.0/train.py}, commit 3ccc44d
}
with the same random seed.

For the calculation of the RMSE in Table~\ref{tab:benchmark} and Table~\ref{tab:grappa}, we subtract the mean predicted energy for each molecule and for each force field since we are only interested in relative energy differences (and the mean is the unique global shift that minimizes the RMSE).
For forces, we report the componentwise RMSE
\begin{align}
    \text{RMSE} &\equiv 
    \sqrt{\frac{1}{3N} \sum_{i=1}^{N} \sum_{j=1}^{3} \left( f_{ij} - f_{ij}^{(\text{ref})} \right)^2}\notag
\end{align}
as it is done in Espaloma~0.3.

\subsubsection{Dataset creation}
Scripts for the creation of the datasets are publicly available.
\footnote{
\url{https://github.com/LeifSeute/grappa-data-creation}
}
We use the Amber ff99SB-ILDN~\cite{amber99sbildn_paper} force field for sampling the states and Psi4~\cite{psi4} for the single point QM calculations.
For sampling radical peptide states, we use a preliminary Grappa model trained on optimization trajectories and torsion scans of radical peptides.
The radical peptides are what we call hydrogen-atom-transfer (HAT) type radicals, that is, they are formed by removing a single hydrogen from a peptide.
For the nonbonded contribution, we use the same Lennard-Jones parameters as for the original peptide and add the partial charge of the hydrogen in the original peptide to the heavy atom it was attached to.



\subsection{MD simulations}
All molecular dynamics simulations are performed with GROMACS version 2023~\cite{gromacs} using the Amber ff99SB-ILDN force field~\cite{amber99sbildn_paper} or Grappa-1.3 with Amber ff99SB-ILDN nonbonded parameters and the TIP3P water model. 2\,fs time steps with LINCS constraints~\cite{hess_p-lincs_2008} on H-bonds were employed. Simulations were run at 300K and 1 bar, maintained by the v-rescale thermostat~\cite{bussi_canonical_2007} and Parrinello-Rahman pressure coupling~\cite{parrinello_polymorphic_1981}. We apply a Coulomb and Lennard-Jones cutoff of 1.0\.nm. After system preparation, energy minimization, NVT and NPT equilibration simulations are conducted.
The code for reproducing the simulations is publicly available.
\footnote{
\url{https://github.com/LeifSeute/validate-grappa}.
}


\subsubsection{Ubiquitin simulations}\label{sec:simulation_stability}
The simulations of ubiquitin (PDB ID: 1UBQ) are evaluated by calculating a moving average over 0.05\,ns of the C-alpha RMSD to the initial structure (Figure~\ref{fig:stability_rmsd}b).
We also calculate the statistics of the C-alpha RMSD as function of time difference between two states.
To this end, we sample 1,000 states from the trajectory for each time difference and calculate the mean C-alpha RMSD between the states.
For the simulation of STMV, we use the same structure files as~\cite{allegro}. We use partial charges and Lennard-Jones parameters of Amber ff99SB-ILDN~\cite{amber99sbildn_paper} for the proteins and of RNA.OL3~\cite{rna_ol3} for RNA.

\subsubsection{STMV virus}

For the simulation of STMV referred to in section~\ref{sec:efficiency}, we use nonbonded parameters from Amber ff99SB and RNA.OL3 repsectively.

\begin{figure}[h]
    \centering
    \includegraphics[width=\columnwidth]{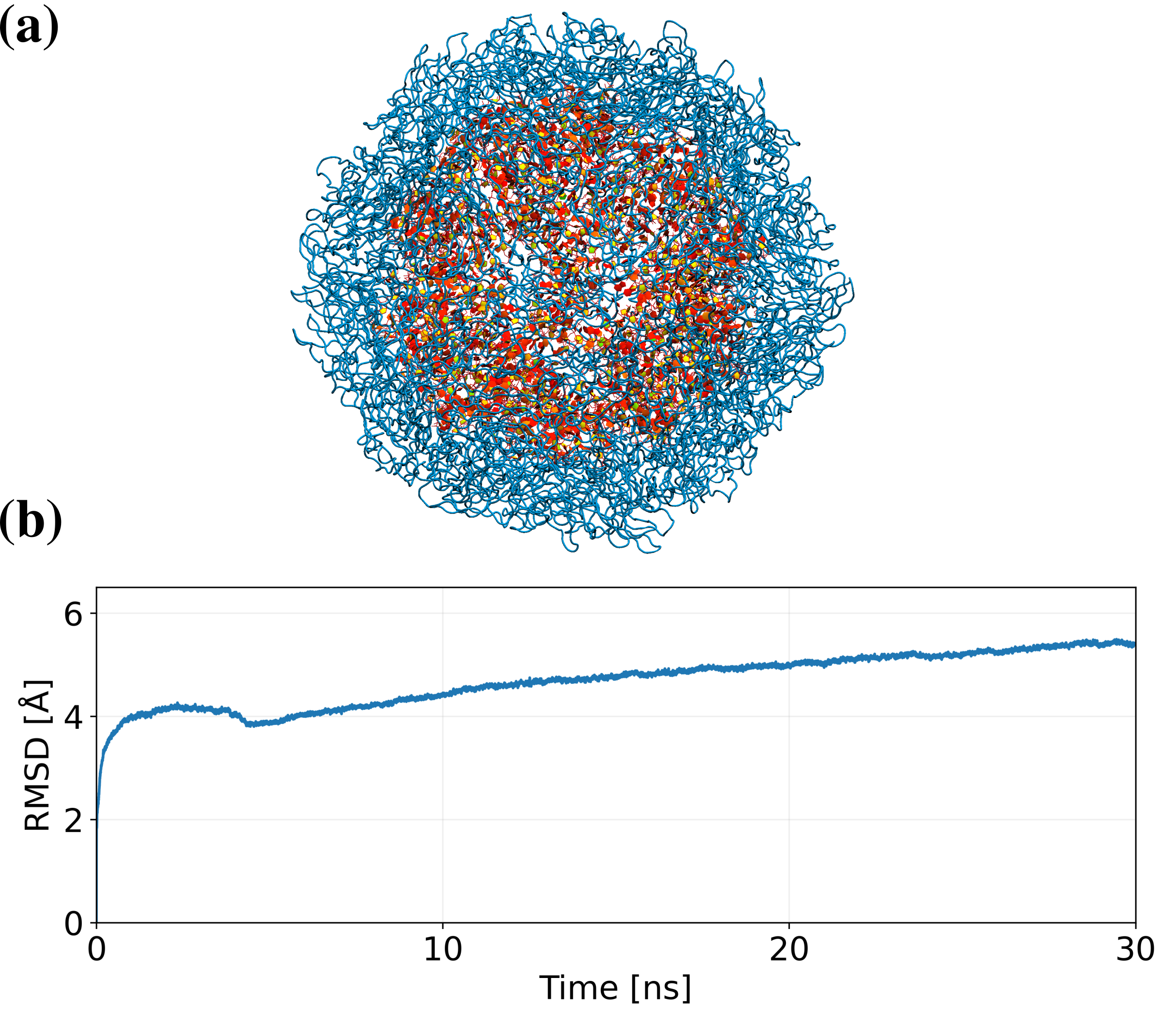}
    \caption{
    \textbf{(a)} The virus STMV, containing proteins (blue) and RNA (red).
    \textbf{(b)} RMSD of protein C-alpha and RNA carbon atoms of the virus in solution during a 30\,ns MD simulation with Grappa.
    }
    \label{fig:virus}
\end{figure}

Fig.~\ref{fig:virus}~(b) shows the RMSD of protein C-alpha and RNA carbon atoms during 30\,ns of MD simulation with Grappa parameters for protein and RNA, which indicates that Grappa keeps the virus stable.
For large scale conformational changes or artificial viral capsid disruption, higher RMSD values would be expected.

\subsection{Learning curve}

\begin{figure*}[h!]
    \centering
    \includegraphics[width=0.8\textwidth]{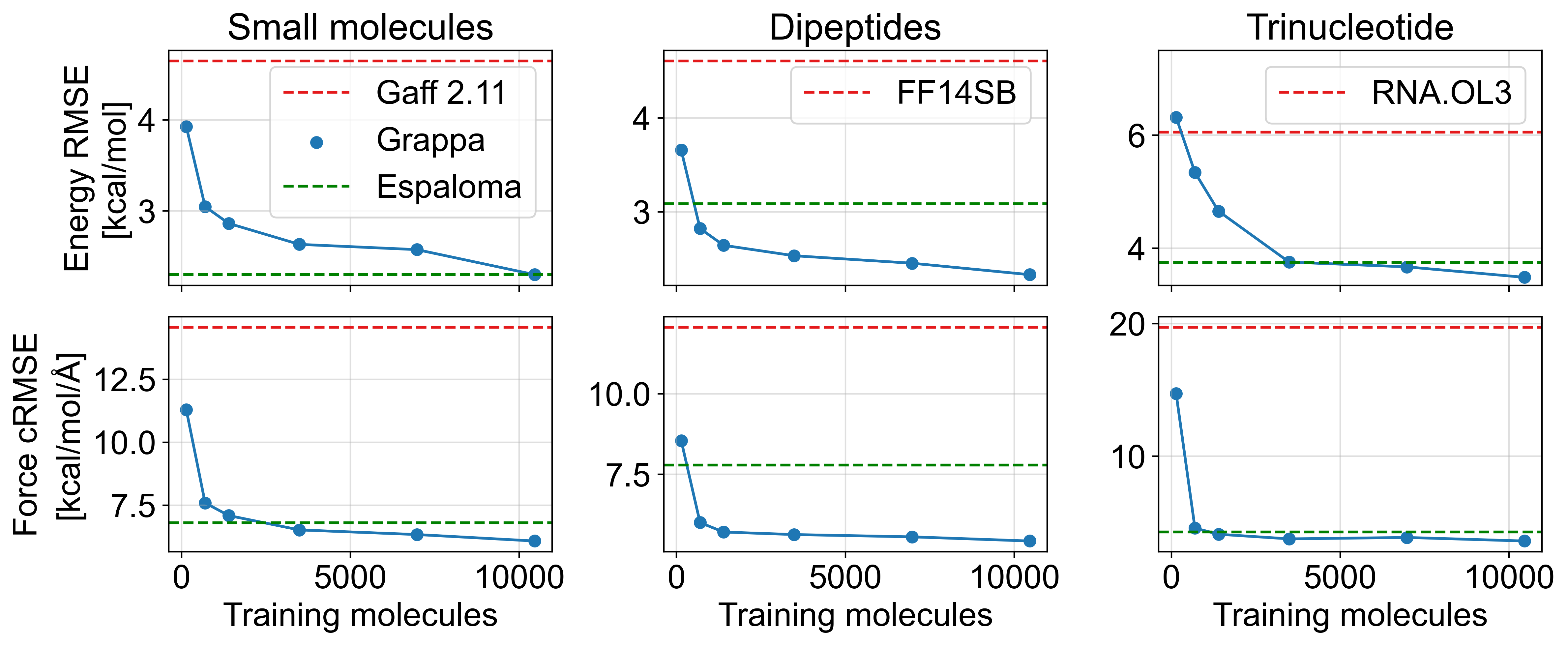}
    \caption{
        Test RMSE of Grappa (blue), trained on a fraction of the Espaloma-0.3 train dataset, compared with the RMSE of Espaloma-0.3 (green) trained on the full dataset and the RMSE of established force fields.
    }
    \label{fig:learning_curve}
\end{figure*}

We provide a learning curve for Grappa, trained on subsets of the Espaloma dataset, in Fig.~\ref{fig:learning_curve}.

\subsection{Grappa's MM parameters}

\begin{figure*}[h!]
    \centering
    \includegraphics[width=0.95\textwidth]{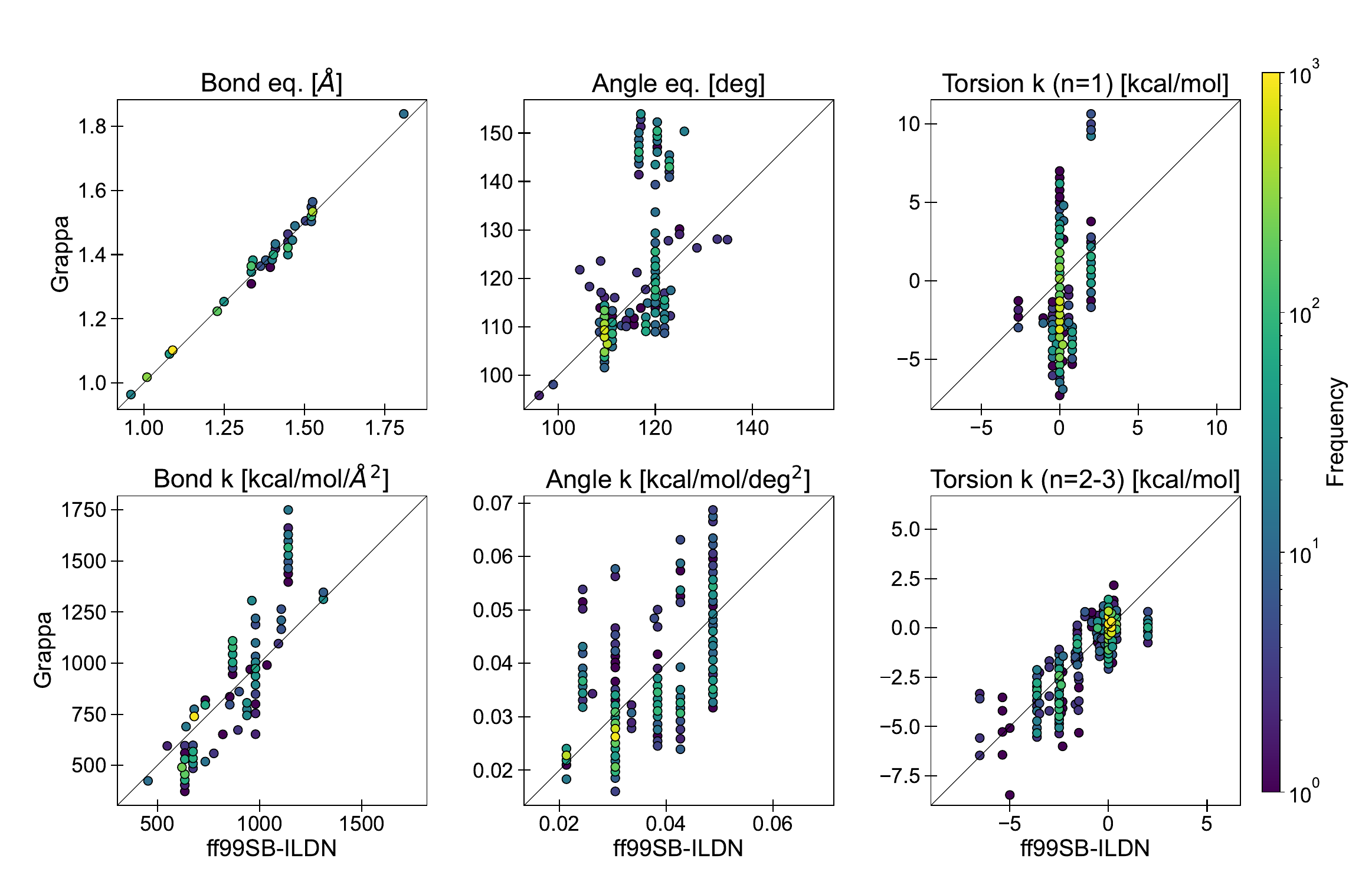}
    \caption{
        MM parameters for the protein ubiquitin, predicted by Grappa-1.3 and ff99SB-ILDN.
        The comparison suggests that Grappa predicts more continuous parameter sets than the tabulated force field ff99SB-ILDN.
    }
    \label{fig:parameter_comparison}
\end{figure*}

In Fig.~\ref{fig:parameter_comparison}, the MM parameters predicted by Grappa and ff99-SBILDN for the protein ubiquitin are visualized.

\subsection{Force accuracy}

\begin{figure*}[h!]
    \centering
    \includegraphics[width=0.95\textwidth]{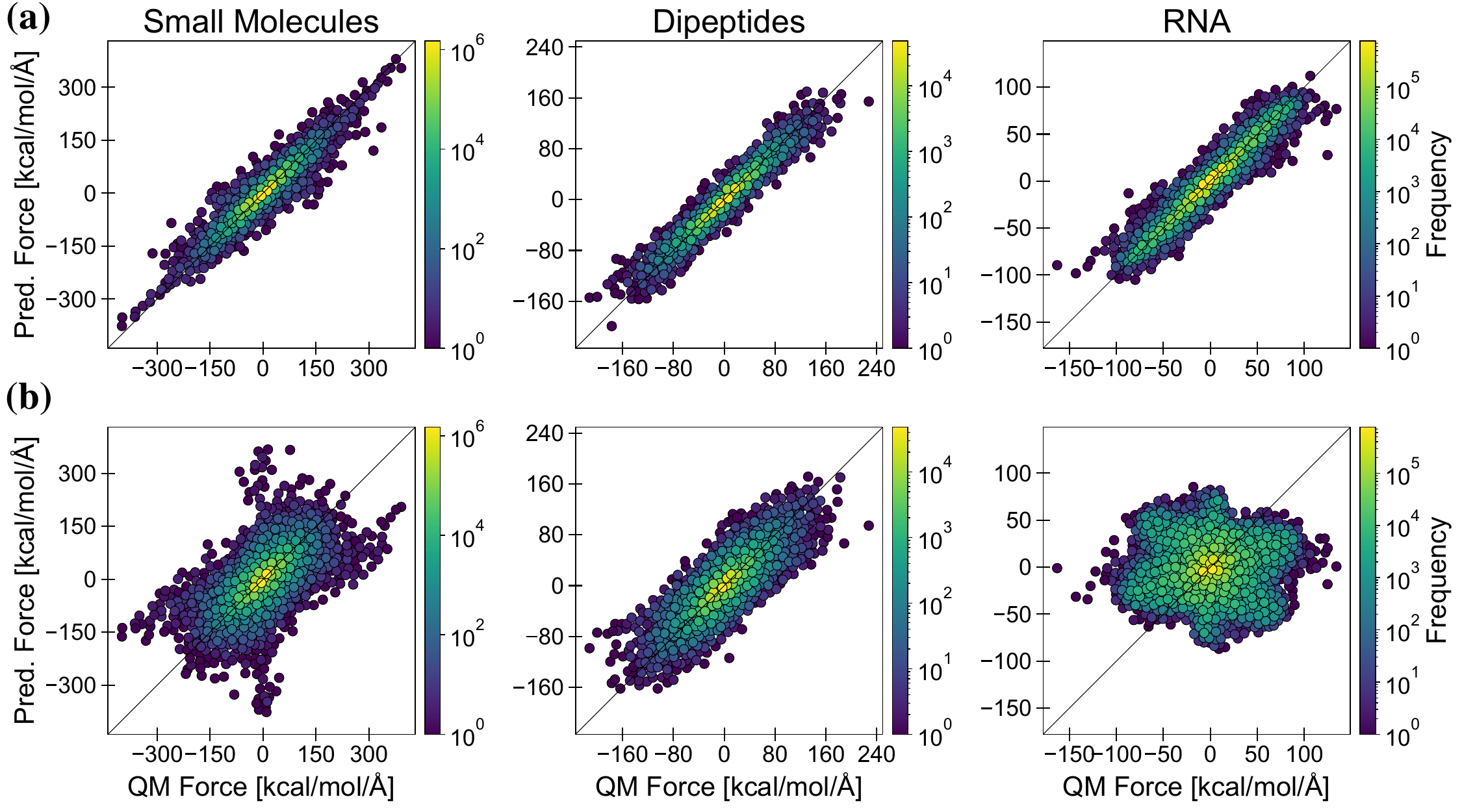}
    \caption{
        Comparison of force predictions of \textbf{(a)} \textbf{Grappa-1.3} and the established force fields \textbf{(b)} \textbf{Gaff-2.11}, \textbf{ff99SB-ILDN} and \textbf{RNA.OL3} for test molecules from Espaloma's SPICE-Pubchem, SPICE-Dipeptide and RNA-Trinucleotide datasets.
        For SPICE-PubChem, five molecules were filtered out because they had unphysically high gradient magnitudes of more than 400 kcal/mol/Å.
    }
    \label{fig:gradient_scatter}
\end{figure*}

In analogy to Fig.~\ref{fig:scatter_and_pca}, Grappa's accuracy for predicted force components is visualized in Fig.~\ref{fig:gradient_scatter}.

\subsection{Extensive tables}

We provide tables for with errors for the full Grappa-1.3 and Espaloma datasets in Tab.~\ref{tab:full_benchmark} and~\ref{tab:full_grappa}.

\onecolumn


\clearpage 

\begin{table*}[h!]
    
\newcommand{\widthbetweentype}{7pt}

\begin{tabular*}{\textwidth}{l c c l c c c c c c}
        \hline
        \hline
        \multirow{2}{*}{Dataset} & \multirow{2}{*}{Test Mols} & \multirow{2}{*}{Confs} & & \multirow{2}{*}{Grappa} & \multirow{2}{*}{Espaloma} & \multirow{2}{*}{Gaff-2.11} & ff14SB, & Mean\\
        & & & & & & & RNA.OL3 & Predictor\\
        \hline
        \multicolumn{8}{l}{\vspace{\widthbetweentype}} \\[-1em]
        \multicolumn{8}{l}{\small{BOLTZMANN SAMPLED}} \\\hline
        \multirow{2}{*}{SPICE-Pubchem} & \multirow{2}{*}{1411} & \multirow{2}{*}{60853} & \textit{Energy} & \textbf{2.3} $\pm$ 0.1 & \textbf{2.3} $\pm$ 0.1 & 4.6 &  & 18.4\\
                                           &                       &                         & \textit{Force}  & \textbf{6.1} $\pm$ 0.3 & 6.8 $\pm$ 0.1 & 14.6 &  & 23.4\\
        \hline
        \multirow{2}{*}{SPICE-DES-Monomers} & \multirow{2}{*}{39} & \multirow{2}{*}{2032} & \textit{Energy} & \textbf{1.3} $\pm$ 0.1 & 1.4 $\pm$ 0.3 & 2.5 &  & 8.2\\
                                           &                       &                         & \textit{Force}  & \textbf{5.2} $\pm$ 0.2 & 5.9 $\pm$ 0.5 & 11.1 &  & 21.3\\
        \hline
        \multirow{2}{*}{SPICE-Dipeptide} & \multirow{2}{*}{67} & \multirow{2}{*}{2592} & \textit{Energy} & \textbf{2.3} $\pm$ 0.1 & 3.1 $\pm$ 0.1 & 4.5 & 4.6 & 18.7\\
                                           &                       &                         & \textit{Force}  & \textbf{5.4} $\pm$ 0.1 & 7.8 $\pm$ 0.2 & 12.9 & 12.1 & 21.6\\
        \hline
        \multirow{2}{*}{RNA-Diverse} & \multirow{2}{*}{6} & \multirow{2}{*}{357} & \textit{Energy} & \textbf{3.3} $\pm$ 0.2 & 4.2 $\pm$ 0.3 & 6.5 & 6.0 & 5.4\\
                                           &                       &                         & \textit{Force}  & \textbf{3.7} $\pm$ 0.04 & 4.4 $\pm$ 0.1 & 16.7 & 19.4 & 17.1\\
        \hline
        \multirow{2}{*}{RNA-Trinucleotide} & \multirow{2}{*}{64} & \multirow{2}{*}{35811} & \textit{Energy} & \textbf{3.5} $\pm$ 0.1 & 3.8 $\pm$ 0.2 & 5.9 & 6.1 & 5.3\\
                                           &                       &                         & \textit{Force}  & \textbf{3.6} $\pm$ 0.01 & 4.3 $\pm$ 0.1 & 17.1 & 19.7 & 17.7\\
        \hline
        \multicolumn{8}{l}{\vspace{\widthbetweentype}} \\[-1em]
        \multicolumn{8}{l}{\small{TORSION SCAN}} \\\hline
        \multirow{2}{*}{Gen2-Torsion} & \multirow{2}{*}{131} & \multirow{2}{*}{21890} & \textit{Energy} & 1.7 $\pm$ 0.2 & \textbf{1.6} $\pm$ 0.3 & 2.7 &  & 4.7\\
                                           &                       &                         & \textit{Force}  & \textbf{4.0} $\pm$ 0.4 & 4.7 $\pm$ 0.6 & 9.4 &  & 5.5\\
        \hline
        \multirow{2}{*}{Protein-Torsion} & \multirow{2}{*}{9} & \multirow{2}{*}{6624} & \textit{Energy} & 2.2 $\pm$ 0.4 & \textbf{1.9} $\pm$ 0.2 & 3.0 &  & 3.5\\
                                           &                       &                         & \textit{Force}  & 3.8 $\pm$ 0.5 & \textbf{3.5} $\pm$ 0.3 & 9.7 &  & 5.1\\
        \hline
        \multicolumn{8}{l}{\vspace{\widthbetweentype}} \\[-1em]
        \multicolumn{8}{l}{\small{OPTIMIZATION}} \\\hline
        \multirow{2}{*}{Gen2-Opt} & \multirow{2}{*}{154} & \multirow{2}{*}{40055} & \textit{Energy} & 1.8 $\pm$ 0.2 & \textbf{1.7} $\pm$ 0.5 & 3.0 &  & 3.9\\
                                           &                       &                         & \textit{Force}  & \textbf{3.8} $\pm$ 0.2 & 4.5 $\pm$ 0.8 & 9.7 &  & 5.1\\
        \hline
        \multirow{2}{*}{Pepconf-Opt} & \multirow{2}{*}{55} & \multirow{2}{*}{14884} & \textit{Energy} & 3.2 $\pm$ 0.3 & \textbf{2.8} $\pm$ 0.3 & 5.1 & 4.1 & 6.3\\
                                           &                       &                         & \textit{Force}  & \textbf{3.6} $\pm$ 0.2 & 4.0 $\pm$ 0.4 & 10.2 & 10.2 & 5.3\\
        \hline
        \hline
        \hline
\end{tabular*}

    \caption{
        Accuracy of Grappa, Espaloma~0.3~\cite{espaloma0.3_paper} and established MM force fields on test molecules from the Espaloma dataset.
        We report the RMSE of molwise-centered energies in kcal/mol and the componentwise RMSE of forces in kcal/mol/Å.
        Gaff-2.11 \cite{gaff_paper} is a general-purpose force field, ff14SB \cite{amber14} is an established protein force field and RNA.OL3 \cite{rna_ol3} is specialized to RNA.
        As in Espaloma, we bootstrap the set of test molecules 1000 times and report mean and standard deviation of the RMSEs.
        }
    \label{tab:full_benchmark}
\end{table*}

\begin{table*}[h!]
    
\newcommand{\widthbetweentype}{7pt}

\begin{tabular}{l c c l c  c c c c c}

\hline
\hline
\multirow{2}{*}{Dataset} & \multirow{2}{*}{Test Mols} & \multirow{2}{*}{Confs} & & \multirow{2}{*}{Grappa} & \multirow{2}{*}{Espaloma} & \multirow{2}{*}{Gaff-2.11} & \multirow{2}{*}{ff99SB-ILDN} & Mean\\
& & & &&&& & Predictor \\
\hline
\multirow{2}{*}{SPICE-PubChem} & \multirow{2}{*}{1411} & \multirow{2}{*}{60853} & \textit{Energy} & \textbf{2.3} $\pm$ 0.1 & \textbf{2.3} $\pm$ 0.1 & 4.6 $\pm$ 0.1 &  & 18.8 \\
                                 & & & \textit{{Force}}  & \textbf{6.1} $\pm$ 0.3 & 6.8 $\pm$ 0.1 & 14.6 $\pm$ 0.3 &  & 41.3 \\
\hline
\multirow{2}{*}{SPICE-DES-Monomers} & \multirow{2}{*}{39} & \multirow{2}{*}{2032} & \textit{Energy} & \textbf{1.3} $\pm$ 0.1 & 1.4 $\pm$ 0.3 & 2.5 $\pm$ 0.2 &  & 8.6 \\
                                 & & & \textit{{Force}}  & \textbf{5.3} $\pm$ 0.2 & 5.9 $\pm$ 0.5 & 11.1 $\pm$ 0.6 &  & 37.4 \\
\hline
\multirow{2}{*}{SPICE-Dipeptide} & \multirow{2}{*}{67} & \multirow{2}{*}{2592} & \textit{Energy} & \textbf{2.4} $\pm$ 0.1 & 3.1 $\pm$ 0.1 & 4.5 $\pm$ 0.1 &  & 19.1 \\
                                 & & & \textit{{Force}}  & \textbf{5.4} $\pm$ 0.1 & 7.8 $\pm$ 0.2 & 12.9 $\pm$ 0.3 &  & 38.4 \\
\hline
\multirow{2}{*}{RNA-Diverse} & \multirow{2}{*}{6} & \multirow{2}{*}{357} & \textit{Energy} & \textbf{3.2} $\pm$ 0.2 & 4.2 $\pm$ 0.3 & 6.5 $\pm$ 0.1 &  & 6.1 \\
                                 & & & \textit{{Force}}  & \textbf{3.7} $\pm$ 0.0 & 4.4 $\pm$ 0.1 & 16.8 $\pm$ 0.1 &  & 30.2 \\
\hline
\multirow{2}{*}{RNA-Trinucleotide} & \multirow{2}{*}{64} & \multirow{2}{*}{23811} & \textit{Energy} & \textbf{3.5} $\pm$ 0.0 & 3.8 $\pm$ 0.2 & 6.0 $\pm$ 0.1 &  & 6.1 \\
                                 & & & \textit{{Force}}  & \textbf{3.6} $\pm$ 0.0 & 4.3 $\pm$ 0.1 & 17.0 $\pm$ 0.0 &  & 31.2 \\
\hline
\multirow{2}{*}{Dipeptides-300K} & \multirow{2}{*}{72} & \multirow{2}{*}{3600} & \textit{Energy} & \textbf{2.6} $\pm$ 0.1 &  & 4.5 $\pm$ 0.1 & 4.1 $\pm$ 0.1 & 7.8 \\
                                 & & & \textit{{Force}}  & \textbf{5.9} $\pm$ 0.1 &  & 10.9 $\pm$ 0.3 & 11.8 $\pm$ 0.1 & 43.7 \\
\hline
\multirow{2}{*}{Dipeptides-1000K} & \multirow{2}{*}{72} & \multirow{2}{*}{2160} & \textit{Energy} & \textbf{5.4} $\pm$ 0.1 &  & 8.6 $\pm$ 0.2 & 8.5 $\pm$ 0.2 & 18.9 \\
                                 & & & \textit{{Force}}  & \textbf{11.6} $\pm$ 0.1 &  & 16.1 $\pm$ 0.2 & 17.8 $\pm$ 0.1 & 74.3 \\
\hline
\multirow{2}{*}{Non-Capped-Peptides} & \multirow{2}{*}{10} & \multirow{2}{*}{500} & \textit{Energy} & \textbf{2.2} $\pm$ 0.2 &  & 4.2 $\pm$ 0.4 & 4.0 $\pm$ 0.3 & 7.2 \\
                                 & & & \textit{{Force}}  & \textbf{6.1} $\pm$ 0.2 &  & 12.1 $\pm$ 0.7 & 12.6 $\pm$ 0.5 & 45.8 \\
\hline
\multirow{2}{*}{Radical-Dipetides} & \multirow{2}{*}{28} & \multirow{2}{*}{272} & \textit{Energy} & \textbf{3.3} $\pm$ 0.3 &  &  &  & 8.7 \\
                                 & & & \textit{{Force}}  & \textbf{6.8} $\pm$ 0.2 &  &  &  & 41.3 \\
\hline

\multicolumn{8}{l}{\vspace{\widthbetweentype}} \\[-1em]
\multicolumn{8}{l}{\small{OPTIMIZATION}} \\\hline

\multirow{2}{*}{Gen2-Opt} & \multirow{2}{*}{154} & \multirow{2}{*}{29055} & \textit{Energy} & \textbf{1.7} $\pm$ 0.1 & \textbf{1.7} $\pm$ 0.5 & 2.8 $\pm$ 0.2 &  & 4.2 \\
                                 & & & \textit{{Force}}  & \textbf{4.0} $\pm$ 0.2 & 4.5 $\pm$ 0.8 & 9.8 $\pm$ 0.3 &  & 8.7 \\
\hline
\multirow{2}{*}{Pepconf-Opt} & \multirow{2}{*}{55} & \multirow{2}{*}{9084} & \textit{Energy} & \textbf{2.8} $\pm$ 0.2 & \textbf{2.8} $\pm$ 0.3 & 4.7 $\pm$ 0.3 &  & 6.5 \\
                                 & & & \textit{{Force}}  & \textbf{3.7} $\pm$ 0.2 & 4.0 $\pm$ 0.4 & 10.4 $\pm$ 0.3 &  & 9.4 \\
\hline

\multicolumn{8}{l}{\vspace{\widthbetweentype}} \\[-1em]
\multicolumn{8}{l}{\small{TORSION SCAN}} \\\hline

\multirow{2}{*}{Gen2-Torsion} & \multirow{2}{*}{131} & \multirow{2}{*}{19290} & \textit{Energy} & 1.7 $\pm$ 0.1 & \textbf{1.6} $\pm$ 0.3 & 2.6 $\pm$ 0.1 &  & 4.7 \\
                                 & & & \textit{{Force}}  & \textbf{4.2} $\pm$ 0.2 & 4.7 $\pm$ 0.6 & 9.5 $\pm$ 0.4 &  & 8.9 \\
\hline
\multirow{2}{*}{Protein-Torsion} & \multirow{2}{*}{9} & \multirow{2}{*}{6024} & \textit{Energy} & \textbf{1.9} $\pm$ 0.2 & \textbf{1.9} $\pm$ 0.2 & 2.9 $\pm$ 0.2 &  & 3.5 \\
                                 & & & \textit{{Force}}  & 4.2 $\pm$ 0.3 & \textbf{3.5} $\pm$ 0.3 & 9.8 $\pm$ 0.4 &  & 9.0 \\
\hline
\hline
\hline
\end{tabular}
    \caption{
        Accuracy of Grappa-1.3 with nonbonded parameters from the AM1-BCC scheme, Grappa with nonbonded parameters from Amber ff99SB-ILDN, Espaloma~0.3~\cite{espaloma0.3_paper} and established MM force fields on test molecules of the Grappa-1.3 dataset.
        We report the RMSE of molwise-centered energies in kcal/mol and the componentwise RMSE of forces in kcal/mol/Å.
        }
    \label{tab:full_grappa}
\end{table*}




\todo{contribution plots for dipeptides-300K}

\listoftodos

\end{document}